\documentclass[12pt, a4paper]{article}
 
\usepackage{a4}
\usepackage{graphicx} 
\usepackage{amsmath}
\usepackage{amssymb}
\usepackage{multirow}
\usepackage{tikz}
\usetikzlibrary{shapes,arrows}
\usetikzlibrary{arrows}
\tikzstyle{decision} = [diamond, draw, 
    text width=4.5em, text badly centered, node distance=3cm, inner sep=0pt]
\tikzstyle{block} = [rectangle, draw, 
    text width=5em, text centered, rounded corners, minimum height=4em]
\tikzstyle{line} = [draw, -latex']
\tikzstyle{cloud} = [draw, ellipse]

\usepackage{hyperref}
\hypersetup{pdftitle={On Instanton Effects in F-theory}, 
      pdfauthor={Ralph Blumenhagen, Andres Collinucci and Benjamin Jurke}, 
      pdfsubject={String theory model building, Instantons},
      pdfstartview={FitH}, pdfpagelayout={TwoColumnRight},
      bookmarksopen, bookmarksnumbered, bookmarksopenlevel=2}

\newcommand{\beq}{\begin{equation}}  \newcommand{\eeq}{\end{equation}}
\newcommand{\bal}{\begin{aligned}}   \newcommand{\eal}{\end{aligned}}

\def\ov{\overline}

\def\IP{\mathbb{P}}
\def\IZ{\mathbb{Z}}

\def\cO{\mathcal{O}}
\def\cD{\mathcal{D}}

\def\cX{\mathcal{X}}    
\def\cB{\mathcal{B}}    
\def\cZ{\mathcal{Z}}    
\def\cE{\mathcal{E}}    
\def\cM{\mathcal{M}}    
\def\cC{\mathcal{C}}    
\def\cA{\mathcal{A}}    

\def\fh{\mathfrak{h}}

\def\ce{\mathrel{\mathop:}=}  

\def\uspc{${}^\big.$}   
\def\lspc{${}_\big.$}

\def\fto{\longrightarrow}
\def\injto{\lhook\joinrel\relbar\!\!\:\joinrel\rightarrow}
\def\surjto{\relbar\joinrel\twoheadrightarrow}
\def\QdPsix{\smash{Q^{dP_6^2}}}   
\def\UpQdPsix{\smash{\tilde Q^{dP_6}}}

\begin{document}

\baselineskip=14pt

\vspace*{-1.5cm}
\begin{flushright}    
  {\small
  MPP-2010-11 \\
  LMU-ASC 05/10
  }
\end{flushright}

\vspace{2cm}
\begin{center}        
  {\LARGE
  On Instanton Effects in F-theory  
  }
\end{center}

\vspace{0.75cm}
\begin{center}        
  Ralph Blumenhagen$^{1}$, Andr\'es Collinucci$^{2}$ and Benjamin Jurke$^{1}$
\end{center}

\vspace{0.15cm}
\begin{center}        
  \emph{$^{1}$ Max-Planck-Institut f\"ur Physik, F\"ohringer Ring 6, \\ 
               80805 M\"unchen, Germany}
               \\[0.15cm] 
  \emph{$^{2}$ Arnold Sommerfeld Center for Theoretical Physics,\\ 
               LMU, Theresienstr.~37, 80333 M\"unchen, Germany} 
\end{center} 

\vspace{2cm}


\begin{abstract}
We revisit the issue of M5-brane instanton corrections to the superpotential in F-theory compactifications on elliptically fibered Calabi-Yau fourfolds. Elaborating on concrete geometries, we compare the instanton zero modes for non-perturbative F-theory models with the zero modes in their perturbative Sen limit. The fermionic matter zero modes localized on the intersection of the instanton with the space-time filling D7-branes show up in a geometric way in F-theory. Methods for their computation are developed and, not surprisingly, exceptional gauge group structures do appear. Finally, quite intriguing geometrical aspects of the one-loop determinant are discussed. 
\end{abstract}

\clearpage

\tableofcontents

\newpage


\section{Introduction}

Recently it was realized that F-theory compactifications on singular Calabi-Yau fourfolds provide an arena for four-dimensional ${\cal N}=1$ supersymmetric GUT like string models, where most of the obstacles of heterotic and perturbative D-brane models can be overcome (see for instance \cite{Donagi:2008ca, Beasley:2008dc, Beasley:2008kw, Donagi:2008kj,Hayashi:2009ge, Andreas:2009uf, Donagi:2009ra, Marsano:2009ym, Blumenhagen:2009yv,Grimm:2009yu} or for a more formal respectively phenomenological review \cite{Denef:2008wq} and \cite{Heckman:2010bq}). First by realizing the GUT symmetry on a 7-brane wrapping a shrinkable four-cycle of del Pezzo type, the size of the gravitational coupling can be decoupled from the gauge coupling, i.e.~there exists a mechanism for explaining the high scale hierarchy between the GUT and the Planck scale. Moreover, in contrast to perturbative orientifolds with only D7-branes, all $SU(5)$ Yukawa couplings are naturally present in F-theory and are generically of the same order of magnitude. Recall that in perturbative Type IIB constructions the ${\bf 10\,10\,5_H}$ Yukawa coupling is absent \cite{Blumenhagen:2001te} and can only be generated by Euclidean D3-brane instantons (also called E3-brane) with the right number of fermionic zero modes \cite{Blumenhagen:2007zk}.

Such D-brane instanton effects in Type II orientifolds were one of the main areas of research in string theory during the last three years \cite{Blumenhagen:2006xt,Ibanez:2006da,Argurio:2007vqa,Ibanez:2007rs} and it has led to a detailed understanding of their zero mode structure and of their effects on the various terms in the effective four-dimensional supergravity theory (see \cite{Blumenhagen:2009qh} for a review). Clearly, for perturbatively absent couplings these instanton effects can become dominant and their understanding is important for the dynamics of the whole system. Since the holomorphic superpotential enjoys the very strong stringy non-renormalization theorem that it receives no string loop corrections at all, one can easily face the aforementioned situation. 

Such corrections are very important in particular for the superpotential of the scalar moduli fields, as these non-perturbative effects can lead to either a disastrous destabilization or a welcomed stabilization of the moduli controlling the action of the instanton. More concretely, it is known that the Type IIB tree-level superpotential only depends on the complex structure moduli, the dilaton and the brane moduli, but not on the K\"ahler moduli. The first correction depending on the K\"ahler moduli therefore comes from E3-brane instantons. It is precisely this structure that is at the core of the KKLT \cite{Kachru:2003aw} and LARGE-volume \cite{Balasubramanian:2005zx} scenario of Type IIB moduli stabilization.

Clearly, analogous effects are also present in non-perturbative F-theory compactifications and need to be understood. While the  tree-level F-theory superpotential has already  been the subject of some recent activities \cite{Alim:2009bx,Grimm:2009ef,Grimm:2009sy}, here we will focus on instanton generated corrections. We consider F-theory on a Calabi-Yau fourfold ${\cal Z}_4$, elliptically fibered over a threefold base ${\cal B}_3 = X_3/\sigma$, which can be regarded as the orientifold quotient of a Calabi-Yau threefold $X_3$. One can define  F-theory on such a fourfold as M-theory on the same fourfold in the limit of vanishing size of the elliptic fiber. It has been argued in \cite{Witten:1996bn} that the M-theory description of the instantons relevant for correcting the ${\cal N}=1$ superpotential are the M5-brane instantons introduced in \cite{Becker:1995kb} (see also \cite{Robbins:2004hx}). If these M5-branes wrap the elliptic fiber entirely, they descend to E3-brane instantons in Type IIB. As derived in \cite{Witten:1996bn}, a necessary condition for such an instanton, wrapping a divisor threefold ${\cal M}_3\subset {\cal Z}_4$, to contribute to the superpotential is that the holomorphic Euler characteristic is equal to one, i.e.~$\chi({\cal M}_3,{\cal O})=1$. Note that such a vertical divisor ${\cal M}_3$ is itself elliptically-fibered over a surface ${\cal E}_{2}\subset {\cal B}_3$. More precisely, the surface ${\cal E}_2 = E_2/\sigma$ can be regarded as the orientifold quotient of the surface $E_2 \subset X_3$ wrapped by the E3-instanton in the Calabi-Yau threefold. 

\begin{figure}[ht]
  \centering
  \includegraphics[width=\textwidth]{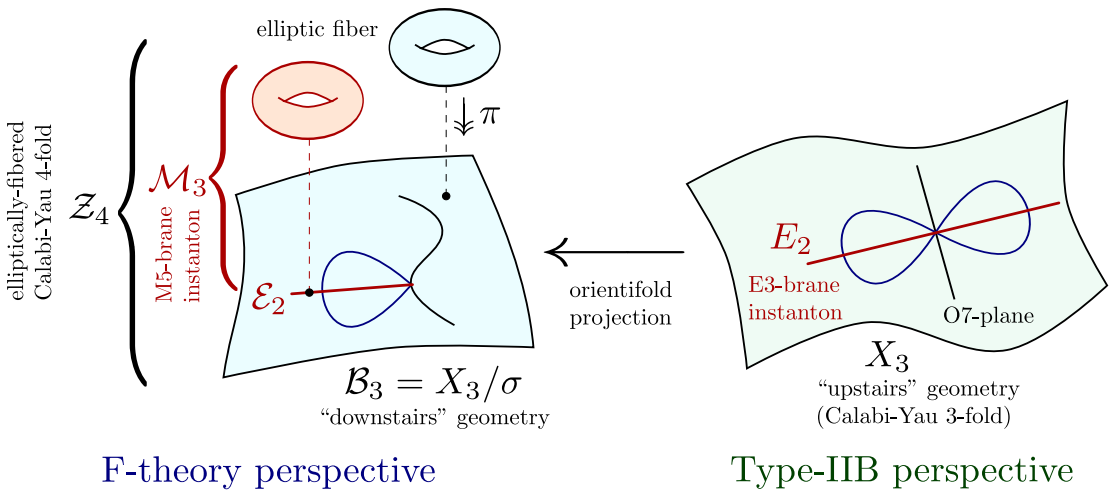}
  \caption{\small Geometric setting}
  \label{fig:notation}
\end{figure}

The situation and our notation are summarized in Figure \ref{fig:notation}. There we have also shown the location of the 7-branes both in F-theory and in the Type IIB orientifold. In F-theory, for a smooth Weierstra\ss\ fibration the elliptic fiber degenerates over a surface in ${\cal B}_3$, which has a cusp singularity. In the weak coupling so-called  Sen limit it splits into an O7-plane and a single orientifold invariant D7-brane with a curve worth of double points on the O7-plane, and a number of pinch points on that curve.

It is the aim of this paper to investigate the zero mode structure of such vertical M5-brane instantons in much more detail\footnote{See \cite{Heckman:2008es, Marsano:2008py, Cvetic:2009mt, Cvetic:2009ah} for recent work in the context of local F-theory GUT models in this direction and see \cite{Berglund:2005dm,Jockers:2009ti} for work on F-theory superpotentials via string dualities.}. This is not a straightforward task since, as opposed to D-brane instantons, the effective theory on the M5-brane is not well understood. It is not a six-dimensional gauge theory like for D-branes but rather a theory with a self-dual two-form. Therefore, this is momentarily not a very practical starting point. We take a back-route and start with the well understood structure of Euclidean E3-branes in Type IIB orientifolds and ask how they uplift to F-theory. For this latter purpose we can use the results of \cite{Collinucci:2008zs, Blumenhagen:2009up, Collinucci:2009uh}, where for certain orientifolds on Calabi-Yau threefolds, the F-theory uplift geometry could be determined explicitly. We will see that this way of studying M5-instantons initially on a case by case basis is very powerful and by generalization allows us to clarify quite a number of issues concerning the zero modes of M5-instantons. For simplicity, we will focus on the case without four-form fluxes.

First, studying the uplift of $O(1)$ E3-brane instantons, we will find the interpretation of the cohomology classes $H^{i,0}({\cal M})$, $i=0,\ldots, 3$ from the Type IIB perspective.\footnote{Part of our results one might be able to anticipate from the earlier work of \cite{Kallosh:2005gs,Bergshoeff:2005yp,Park:2005hj}. We would like to thank Volker Braun for pointing this out to us.}  In M-theory these modes are given by Euclidean M2-branes with two-dimensional boundaries ending on the M5-brane. Second, the uplift of $U(1)$ instantons turns out to be highly non-trivial and depends on whether the M5-brane intersects certain components of the discriminant locus, i.e.~if the elliptically fibered threefold ${\cal M}$ degenerates over certain curves $C$ in the instanton base ${\cal E}$. If there is no degeneration at all, the topology is ${\cal M}=T^2\times {\cal E}$ and we indeed find the additional universal zero modes (denoted as $\ov\tau_{\dot\alpha}$ in \cite{Blumenhagen:2007bn}). If however the $SL(2,\mathbb Z)$ monodromy is generic, then the zero mode $\ov\tau_{\dot\alpha}$ is non-perturbatively lifted. 

From E3-brane instantons in Type IIB orientifolds one expects that beyond the already mentioned zero modes, there are also so-called matter zero modes, which arise from the intersection of the E3-instanton with the D7-branes \cite{Blumenhagen:2006xt, Ibanez:2006da}. We will argue that such zero modes are also present in F-theory and arise from degenerations of the elliptically-fibered M5-brane ${\cal M}$ over curves in the base ${\cal E}$. These degenerations generically extend to higher rank over points in ${\cal E}$, which in this context encode the Yukawa type couplings between one matter field and two instantonic matter zero modes. These latter couplings are important for absorbing the fermionic zero modes. They arise from Euclidean M2-branes wrapping three-chains with  boundaries ending on the M5-brane. 

In generic settings the discriminant locus on the F-theory side splits into a number of smooth components and a singular remainder where the fiber degenerates to non-ADE Kodaira singularity type $I_1$. Matter zero modes arising from the intersection with smooth components  can be handled rather canonically --- it is not necessary to resolve the ADE degeneration of the fiber along the intersection curve to properly count the modes. However, the situation is more involved for a  singular  component, because both the intersection curve between the E3- and D7-brane in Type IIB and the intersection curve with the M5-brane in F-theory both contain singularities. This requires a careful desingularization on both sides in order to compare the relevant quantities for charged matter zero mode counting. 

The non-perturbative perspective obtained from this matching reveals some interesting insights: On the F-theory side some of the matter zero modes pair up if one moves away from the perturbative  Sen limit. Most importantly, there are charged zero modes on the F-theory side which are not counted by $\chi(\cM,\cO_\cM)$. This  leads to a strong  sufficient criterion for the existence of destabilizing instanton corrections to the superpotential, which is quite similar to the analogous heterotic result for world-sheet instantons.

Another aspect we will study is a direct relation between the topology of the E3-instanton divisor  and the (resolved) E3-D7 intersection curve $\overline{C}$ on the Type IIB side, to the topology of the M5-brane $\cM$ on the F-theory side. Naively one would expect that suitable involution-even/odd combinations of the cycles in the base $\cE$ combined with the $T^2$ fiber cycles to entirely describe the M5-brane topology, but we will find that some cycles do not directly arise in this fashion. Indeed there are additional three-cycles arising from $\sigma$-odd one-cycles of the curve $\overline{C}$. This ultimately allows to express the entire M5-brane topology (i.e.~Hodge diamond) in terms of the E3-brane topology and the genus of the (resolved) Type IIB E3-D7 intersection curves.
\vskip 2mm
This paper is organized as follows: In section 2, we review known aspects of E3-instanton zero modes in orientifold compactifications, establish some definitions and motivate a relation between E3-instanton zero modes and M5-brane zero modes. In section 3, we focus on identifying the uncharged zero modes of both $O(1)$- and $U(1)$-type E3-brane instantons. By studying a concrete example, we infer the rules for translating these zero modes into the zero modes of their uplifted, M5-brane instanton counterparts. From this, we establish a general dictionary between the cohomology of the M5-brane divisor, and the $\mathbb{Z}_2$-equivariant cohomology of the E3-brane divisor. We find that the $\ov \tau$-mode of a $U(1)$-instanton generically gets non-perturbatively lifted. We then proceed to study an explicit $U(1)$-instanton that avoids this lifting.

In section 4, we proceed to uncover the nature of charged zero modes arising from the intersection of the E3 and the D7-branes present. We first study the modes arising from intersections with stacks of smooth, GUT-like D7 stacks. Guided also by duality to the heterotic string, we propose how this generalizes to genuine F-theory models. We make a novel, concrete proposal for determining the Yukawa couplings that are generated by such modes. We then move on to study the charged zero modes arising from the intersection of the E3-brane with a generic D7-brane. Since such a brane necessarily has a singular shape, this requires a desingularization of the intersection curve. We study the counterpart of these modes in the ``downstairs'', geometry, away from the perturbative Sen limit. We find that, non-perturbatively, some of these charged modes are always lifted. 

We then proceed to complete the study of the cohomology of the M5-brane. We propose a dictionary that accounts for the full Hodge diamond of the M5-brane in terms of the equivariant cohomologies of the E3-brane and of the (desingularized) intersection curve between the E3-brane and a generic D7-brane. Finally, we give an interpretation of the three-cycles of the M5-brane, using the analysis of \cite{Witten:1996hc}. We find, that the charged zero modes and the NSNS B-field both play a non-trivial role in determining possible cancellations of the one-loop determinant of the superpotential.

\section{Instantons in Type IIB and F-theory}

Before moving to F-theory, let us start by collecting a number of well established results on the zero mode structure of Euclidean D3-brane instantons in Type IIB orientifolds.

We are considering the Type IIB superstring compactified on a Calabi-Yau threefold $X$ and take the quotient by an orientifold projection $\Omega \sigma (-1)^{F_L}$, where $\Omega$ denotes the world-sheet parity transformation, $F_L$ the left-moving space-time fermion number and $\sigma$ a holomorphic involution satisfying $\sigma^*(J)= J$ and $\sigma^*(\Omega_3)= -\Omega_3$. Here $\Omega_3$ is the holomorphic $(3,0)$-form on $X$.\footnote{For GUT model building in this framework see \cite{Blumenhagen:2008zz}.}

The fixed point locus of the holomorphic involution $\sigma$ defines a holomorphic four-cycle in $X$ and supports an O7-plane. The induced R-R eight-form tadpole is canceled by introducing D7-branes in the background, which in general wrap different four-cycles. The massless modes in four-dimensions consist of a number of $U(N)$ and $SO(2N)/SP(2N)$ gauge fields with charged matter fields localized on the intersections between pairs of D7-branes. In general these intersections are two-cycles in $X$, so that chirality only appears by turning on additional background gauge fields on the D7-branes.

Now, one can consider the effects Euclidean E3-instantons can have on these models. Such an instantonic brane also wraps a holomorphic four-cycle $E$ on $X$ and is point-like in the four-dimensional space. For deciding which four-dimensional couplings receive corrections from such an instanton, it is important to know its zero mode structure. This has been studied and here we summarize the main findings which are important for us.

First there are zero modes arising from open strings with both ends on the E3-brane. If the four-cycle $E$ is not in a $\sigma$ invariant position, a mirror image $E'$ of the E3-brane has to be introduced. Since this pair of branes carries a unitary Chan-Paton label, these instantons are called $U(1)$ instantons. Since the orientifold projection just maps the $E-E$ open strings to the $E'-E'$ open strings, one gets the set of zero modes in Table \ref{tabledeform}.

\begin{table}[ht]
  \centering
  \begin{tabular}{c|c|c}
    zero modes                                     & statistics   & number \\ 
    \hline\hline
    $X_\mu$                                        & bose         & $1$\uspc \\
    $\theta_\alpha$                                & fermi        & $1$ \\
    $\ov\tau_{\dot\alpha}$                         & fermi        & $1$ \\
    $(w, \gamma_{\alpha}, \ov\gamma_{\dot\alpha})$ & (bose,fermi) & $H^{1,0}(E)$     \\
    $(c, \chi_{\alpha}, \ov\chi_{\dot\alpha} )$    & (bose,fermi) & $H^{2,0}(E)$  \\
  \end{tabular}
  \caption{\small $E-E$ zero modes for $U(1)$ instanton.}
  \label{tabledeform}
\end{table} 
 
The zero modes $X_\mu$, $\theta_\alpha$ and $\ov\tau_{\dot\alpha}$ are also called universal zero modes, as they do not depend on the internal geometry of the four-cycle $E$. The remaining ones can be considered as  Wilson-line and deformation zero modes, i.e.~as Goldstone bosons and Goldstinos of brane deformation moduli. For contributions to the holomorphic superpotential the anti-holomorphic $\ov\tau^{\dot\alpha}$ zero modes have to be removed, which happens if the instanton is placed in an orientifold invariant position. In this case the $\Omega\sigma (-1)^{F_L}$ projection acts non-trivially on the zero modes in Table \ref{tabledeform}. In particular, the Wilson-line and deformation moduli split into $\sigma$ even and odd parts
\beq
  H^{1,0}=H_+^{1,0} \oplus H_-^{1,0}, \qquad
  H^{2,0}=H_+^{2,0} \oplus H_-^{2,0}\; .
\eeq
In the case that on the world-volume the projection acts by anti-symmetrization, one gets an $O(1)$ instanton with the zero modes \cite{Argurio:2007vqa,Argurio:2007qk,Ibanez:2007rs,Blumenhagen:2007bn} shown in Table \ref{tabledeformb}.

\begin{table}[ht]
  \centering
  \begin{tabular}{c|c|c}
    zero modes                    &  statistics   & number \\ 
    \hline \hline
    $(X_\mu,\theta_\alpha)$       & (bose, fermi) & $1$\uspc \\
    $\ov\tau_{\dot\alpha}$        & fermi         & $0$ \\
    $\gamma_{\alpha}$             & fermi         & $H_+^{1,0}(E)$     \\
    $(w, \ov\gamma_{\dot\alpha})$ & (bose, fermi) & $H_-^{1,0}(E)$ \\
    $\chi_{\alpha}$               & fermi         & $H_+^{2,0}(E)$  \\
    $(c , \ov\chi_{\dot\alpha} )$ & (bose, fermi) & $H_-^{2,0}(E)$ 
  \end{tabular}
  \caption{\small $E-E$ zero modes for $O(1)$ instanton.}
  \label{tabledeformb}
\end{table}
 
One can define two chiral indices. First there is the usual holomorphic Euler characteristic of the divisor $E$
\beq
  \chi(E,{\cO}_E)=\sum_{i=0}^2 (-1)^i\, h^i(E,{\cO}_E) =\sum_{i=0}^2 (-1)^i\, h^{i,0}(E) \;  
\eeq     
and second one can define an index taking the $\mathbb Z_2$ action of $\sigma$ into account
\beq
  \chi^\sigma(E,{\cO}_E)=\sum_{i=0}^2 (-1)^i\, \left( h_+^{i,0}(E)-h_-^{i,0}(E)\right) \; . 
\eeq
Looking more closely at  the chirality and statistics of the zero modes listed in Table \ref{tabledeform}, motivates us to define a four vector $\fh=(\fh_0,\fh_1,\fh_2,\fh_3)$ by
\beq
  \fh(E)=\bigl(h^{0,0}_+(E)\,,\, h^{0,0}_-(E) + h^{1,0}_+(E)\, ,\, h^{1,0}_-(E) + h^{2,0}_+(E) \,, \,  h^{2,0}_-(E)\bigr)\; .
\eeq
Obviously, its index satisfies
\beq
  \chi^\sigma(E,{\cO}_E)=\sum_{i=0}^3 (-1)^i\ \fh_i(E)\; .
\eeq

One of the objectives of this paper is to clarify the relation between the description of the just described Euclidean D3-brane instantons in Type IIB orientifolds and their uplift to vertical M5-brane instantons in F-theory, i.e.~more precisely the zero size fiber limit $\rho\to 0$ of vertical M5-brane instantons in M-theory. A vertical divisor ${\cal M}$ is one which can be written as the pre-image of a complex two-dimensional divisor $\cE\subset \cB$, i.e.~${\cal M}=\pi^{-1}(\cE)\subset\cZ$. The divisor $\cE$ is the quotient of the divisor $E$ on which the E3-brane is wrapped by the involution $\sigma$. Concerning the latter, it was argued in \cite{Witten:1996bn} that a necessary condition for such an instanton to contribute to the superpotential is 
\beq
  \chi({\cal M},{\cO}_{\cal M})=\sum_{i=0}^3 (-1)^i\, h^i({\cal M},{\cO}_{\cal M}) = 1\; . 
\eeq
A sufficient condition is that beyond the universal zero modes, there are no additional deformation zero modes, i.e.~$h^{1,0}({\cal M})=h^{2,0}({\cal M})=h^{3,0}({\cal M})=0$. Apparently, for F-theory instantons there are four relevant cohomology classes $h^{i,0}({\cal M})$ and for Type IIB orientifold instantons only three. 

It is well known from Type IIB orientifolds that the so far discussed zero modes are not complete. There also exist fermionic zero modes from the intersection of the instanton with the space-time filling D7-branes. These are the so-called charged matter zero modes and their existence provides further strong constraints for which four-dimensional (charged) couplings can receive instanton corrections. For an E3-brane wrapping a four-cycle $E$ and a stack of D7-branes wrapping a four-cycle $D$ and carrying a $U(N)$ Chan-Paton gauge group, these matter zero modes are localized on the intersection curve $C=E\cap D$. If the D7-branes carry a line bundle $L$ with connection in the diagonal $U(1)\subset U(N)$, the number of such matter zero modes is counted by the cohomology groups
\beq
  H^i\bigl(C, L\otimes K_C^{1/2}\bigr), \quad i\in\{0,1\}\; . 
\eeq
This implies that chiral zero modes are only possible for a non-trivial line bundle $L$.

By uplifting, one also expects the appearance of such matter zero modes for the M5-brane instanton in F-theory. These modes must come from the intersection of the divisor ${\cal M}$ with the components of the discriminant of the elliptically fibered Calabi-Yau fourfold. It is the second aim of this paper to properly describe these matter zero modes in F-theory and give clear conditions in which case an M5-instanton can contribute to which coupling in the (charged part of the) superpotential.

\section{Uncharged zero modes}

The aim of this section is to exemplify the appearance of the uncharged zero modes from Table \ref{tabledeformb} for a concrete example. Most explicit computations in the literature have been carried out for toroidal orientifolds, where conformal field theory techniques are available. Here we will discuss genuine Calabi-Yau geometries so that appropriate methods to compute the relevant line bundle cohomology groups need to be employed. The example we will use is the Calabi-Yau threefold $\IP^4_{11114}[8]$ where the orientifold involution acts as a reflection of the homogeneous coordinate of degree four. The F-theory uplift of this simple orientifold has been discussed in quite some detail in \cite{Collinucci:2008pf} and is given by the Weierstra\ss\ fibration over the smooth threefold base $\IP^3$. For completeness, let us first summarize some geometric facts about the Type IIB geometry.

\subsection{The orientifold geometry}\label{sec_31}

Let us consider the Type IIB superstring compactified on the Calabi-Yau threefold $X=\IP^4_{11114}[8]_{(149,1)}$. We denote by $(u_1,\dots,u_4,\xi)$ the homogeneous coordinates of the weighted projective space. As indicated, the threefold hypersurface has $h^{1,1}=1$ K\"ahler moduli and $h^{1,2}=149$ complex structure moduli, which leads to an Euler characteristic of $\chi(X)=-296$. For properly defining the weighted projective space, one has to remove the point $u_1=u_2=u_3=u_4=\xi=0$. In more sophisticated terms, the information about this removed set is given by the Stanley-Reisner ideal
\beq
  \mathrm{SR}(X)= \langle u_1 u_2 u_3 u_4 \xi \rangle\; .
\eeq
In order for $X$ to be invariant under the orientifold involution $\Omega \sigma (-1)^{F_L}$ where
\beq
  \sigma:\xi\mapsto -\xi\; ,
\eeq
the  homogeneous degree-8 polynomial has to be chosen appropriately, i.e.~part of the complex structure moduli are fixed. Let us define the basic divisor in $H_4(X,\IZ)$ by $H=\{u_i=0\}$. The fixed
point locus $\xi=0$ of the orientifold projection defines the O7-plane, which homologically is a divisor $[O7]=4H\in H_4(X,\IZ)$. The triple intersection of $H$ in $X$ is readily found to be $I_{X}=2\, H^3$. Note that the tangent bundle of the Calabi-Yau threefold is defined by the cohomology ${\rm T}_X={\rm Kern}f/{\rm Im}\,g$  of the sequence
\beq
  0\fto {\cal O}_X \stackrel{g}{\fto} \bigoplus_{i=1}^4 {\cal O}_X(1) \oplus {\cal O}_X(4) \stackrel{f}{\fto} {\cal O}_X(8) \fto 0\; .
\eeq

\begin{figure}[ht]
  \begin{center}
    \includegraphics[width=0.5\textwidth]{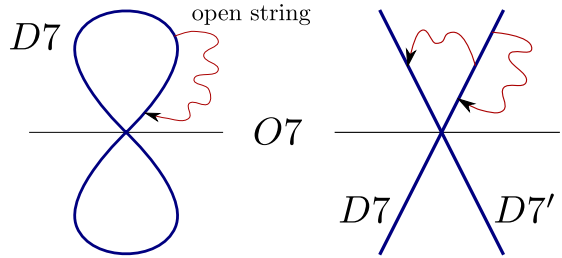}
  \end{center} 
  \caption{\small single $O(1)$ D7-brane - $U(1)$ brane/image-brane pair}
  \label{fig_dseven}
\end{figure}

The orientifold plane induces a tadpole $32 H$, which has to be canceled by additional D7-branes. The intriguing result of Sen \cite{Sen:1996vd,Sen:1997gv} is that in the case that only a single D7-brane is present, it has to have a double intersection with the O7-plane, i.e.~it is not the most generic hypersurface in $32 H$ but has the constrained form
\beq\label{dsevensingle}
  \eta^2 _{n} - \xi^2\, \chi_{2n-8} =0\; .
\eeq
for $n=16$. For later purposes we have left the degree of the constraints as a free parameter. Schematically, such an $O(1)$ brane is shown in the left of Figure \ref{fig_dseven}.

For the special choice $\chi_{2n-8}=\psi^2_{n-4}$ with $n\ge 4$ the single orientifold invariant D7-brane carrying a Chan-Paton gauge group $SO(1)$ splits into a brane/image-brane pair 
\beq\label{dsevenpair}
  \eta^2 _{n} - \xi^2\, \chi_{2n-8} = (\eta _{n} + \xi\, \psi_{n-4})(\eta _{n} - \xi\, \psi_{n-4})=0\; 
\eeq
carrying a gauge group $U(1)$. We  denote the respective four-cycles as $D$ and $D'$. Recombining this pair of D7-branes by giving VEV to massless fields localized on the intersection D7$-$D7' gives back the single brane \eqref{dsevensingle}. However, as pointed out in \cite{Collinucci:2008pf}, there is a subtlety as the D3-brane tadpole contributions of the two brane configurations are not equal. This is reconciled by turning on a non-trivial line bundle on the brane/image-brane configuration
\beq
  {\rm D7}:\  L ={\cal O}\left({\textstyle\frac{n-4}{2}}\right), \qquad\quad 
  {\rm D7'}:\ L'={\cal O}\left(-{\textstyle\frac{n-4}{2}}\right)\; .
\eeq 
For more details we refer the reader to ref.~\cite{Collinucci:2008pf}.

Since it will become important later, here we discuss the moduli spaces of the single brane \eqref{dsevensingle} and brane/image-brane pair \eqref{dsevenpair} in more detail. The moduli space of the single $SO(1)$ brane consists of those deformations which preserve the non-generic form \eqref{dsevensingle}. The number of these deformations has been computed in \cite{Collinucci:2008pf} as
\beq
  \bal
    N_{SO(1)}&=\binom{n+3}{3} + \binom{2n-8+3}{3} - \binom{n-8 +3}{3} -1\\
             &=\frac{4}{3}n^3-8n^2 +\frac{59}{3} n\; .
  \eal
\eeq
This is expected to be the same as the deformations of the brane/image-brane system after brane recombination has been taken into account. 

First there are the transverse deformations of the D7-brane. These can simply be counted as follows:\footnote{A more formal computation will be presented in the next subsection.} For simplicity we write the degree-8 hypersurface constraint in $\IP^4_{11114}$ as $\xi^2=P_8(\vec u)$. Then up to an overall rescaling, the deformations of a generic degree $n$ hypersurface in $X$ are given by all polynomials of degree $n$ modulo the relation $\xi^2=P_8(\vec u)$. The latter allows us to get rid of all terms $\xi^k$ with $k\ge 2$. Therefore, we solely have to count the number of polynomials $P_n(\vec u)$ and $P_{n-4}(\vec u)$ and get
\beq\label{deformmodea}
  N_{{\rm D7-D7}}= \binom{n+3}{3}+ \binom{n-1}{3}-1=\frac{n}{3}(n^2+11)-1\; .
\eeq

In addition one has the brane recombinations localized on the intersection curve $C=D\cap D'$, which naively one might expect to be counted by $H^*(C,L^2\otimes \smash{K_C^{1/2}})$. However, the situation is a bit more subtle, as one has to take the orientifold projection into account. Open strings stretched between the D7 and its mirror D7' brane get symmetrized respectively anti-symmetrized. Clearly, the anti-symmetric modes are completely projected out so that one is only interested in the symmetrized modes, which are counted by $\smash{H^*_-(C,L^2\otimes K_C^{1/2})}$. The chiral index of these modes can be computed via \cite{Douglas:2006xy,Blumenhagen:2008zz}
\beq\label{chiralantiindex}
  \bal
    \chi_-({\rm D7,D7'})&=\frac{1}{2}\left( \int_{X} D\wedge D'\wedge c_1(L^2) -2\,  \int_{X} D\wedge O7\wedge c_1(L) \right) \\
                  &=n^2(n-4) - 4\, n (n-4)=n (n-4)^2\; .
  \eal
\eeq

The problem of really computing the individual cohomology groups has been solved in \cite{Blumenhagen:2006wj}, to which we refer for more details\footnote{The symmetrized modes arise entirely from that part of the intersection curve $C$, which does not lie on the orientifold plane. This means that it is localized on the curve 
\beq
  C'=D\cap D'-D\cap O7\; ,
\eeq
more precisely on the $\mathbb Z_2$ quotient of this curve $C'/\sigma$. However, this quotient is singular as $C'$ still does intersect the orientifold plane in $R = D\cap (D- O7)\cap O7$ ramification points. Taking this properly into account amounts to the following number of symmetric modes
\beq
  \bal
    H^i_-(C,L^2\otimes K_C^{1/2})&= H^{i} (  C'/ \sigma\ , \,  \tilde L   |_{ C' / \sigma}) \\
    &{\rm with}\ \ c_1(\tilde L)|_{ C' / \sigma} = c_1 ( L \otimes K_{D7}^{1/2} ) |_{C'} - \frac {R}{2}\; .
  \eal
\eeq}.
What we can however quite easily compute is the total cohomology class $H^1(C,L^2\otimes K_C^{1/2})$, which via  Serre duality is related to $H^0(C, {\cal O}_C(4))$. For $n>4$ this is simply the number of global sections of ${\cal O}_{X}(4)$, which gives $H^0(C, {\cal O}_C(4))=36$. As shown in \cite{Blumenhagen:2006wj}, the modes localized on the O7-plane, i.e.~on the curve $C_{O7}=O7\cap D$, contribute only to $H^i_+(C,L^2\otimes K_C^{1/2})$. Here again we can easily determine $H^1(C_{O7},L\otimes K_{C_{O7}}^{1/2})$, which is related by Serre duality to $H^0(C_{O7}, {\cal O}_{C_{O7}}(4))=35$. Therefore, we can already conclude that $\smash{H^1_-(C,L^2\otimes K_C^{1/2})}$ can only be zero or one. If it were zero, then the brane/image-brane pair could not recombine, as the D-term constraint of the $U(1)$ gauge symmetry on the D7$-$D7' pair could not be satisfied. Since the recombined cycle exists, we must have $\smash{H^1_-(C,L^2\otimes K_C^{1/2})=1}$ so that
\beq\label{recombmodesa}
  H^0_-(C,L^2\otimes K_C^{1/2})=n (n-4)^2  +1, \qquad
  H^1_-(C,L^2\otimes K_C^{1/2})=1\; . 
\eeq
The total number of recombination modes is $N^{\rm S}_{\rm D7-D7'}=n (n-4)^2+2$. Adding up all the zero modes from the eqs. \eqref{deformmodea} and \eqref{recombmodesa} we find the relation
\beq
  N_{SO(1)}=N_{\rm D7-D7}+N^{\rm S}_{\rm D7-D7'}-1\; ,
\eeq
which is precisely what one expects for the dimension of the moduli space in a brane recombination process. The extra $(-1)$ indicates that the $U(1)$ D-term constraint lifts one of the D7$-$D7' modes.

\subsection[\texorpdfstring{$O(1)$}{O(1)} E3-brane instantons]{\boldmath$O(1)$ E3-brane instantons}

\begin{figure}[ht]
  \begin{center} 
    \includegraphics[width=0.5\textwidth]{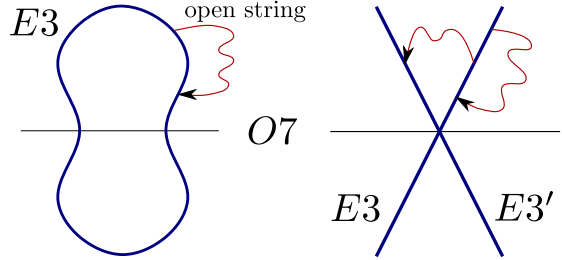}
  \end{center} 
  \caption{\small single $O(1)$ E3-brane - $U(1)$ brane/image-brane pair}
        \label{fig_ethree}
\end{figure}

Now we consider a Euclidean E3-brane wrapping a divisor $E\in H_4(X,\IZ)$, which is also invariant under the orientifold projection. Note that as opposed to the just discussed D7-brane case, there is no reason why the E3-brane cannot wrap the most generic hypersurface in its homology class. This is schematically shown in Figure \ref{fig_ethree}.

Now we will compute the uncharged instanton zero modes for such a Euclidean E3-brane. Let us consider the divisors $E_n$ defined by a generic hypersurface constraint of degree $n$. The intersection form $I_{X}$ on the Calabi-Yau threefold induces the intersection form $I_{E_n}=2n H^2$ on the divisor $E_n$. Moreover, the tangent bundle of $E_n$ is defined via the sequence
\beq
  0\fto {\cal O}_{E_n} \fto \bigoplus_{i=1}^4 {\cal O}_{E_n}(1) \oplus {\cal O}_{E_n}(4) \fto {\cal O}_{E_n}(8) \oplus {\cal O}_{E_n}(n) \fto 0\; .
\eeq
Using that the total Chern class of ${\rm T}_{E_n}$ is given by
\beq
  c({\rm T}_{E_n})=\frac{(1+H)^4\, (1+4H)}{(1+8H)\, (1+nH) }\bigg|_{E_n} =1-nH + (n^2+22)\, H^2  \; ,
\eeq
it is straightforward to compute the holomorphic Euler characteristic 
\beq\label{eulerchar}
  \bal
    \chi(E_n,{\cO}_{E_n})&= \frac{1}{12}\int_{E_n} \left( c_1^2(E_n) + c_2(E_n)\right)\\
     &= \frac{n}{3}\left( n^2 +11 \right) \; .
  \eal
\eeq
For computing the individual cohomology classes $h^i(E_n,{\cO})$ we have to work a little more. For this purpose we first employ the Koszul sequence
\beq
  0 \fto {\cO}_X(-n) \injto {\cO}_{X} \surjto {\cO}_{E_n} \fto 0
\eeq
to relate the cohomology on $E_n$ to the one on the Calabi-Yau threefold $X$. Here we have $h^*({\cO}_{X})=(1,0,0,1)$ and for ${\cO}_X(-n)$ we are using a second Koszul sequence
\beq
  0 \fto  {\cO}_A(-n-8) \injto {\cO}_A(-n) \surjto {\cO}_X(-n) \fto 0
\eeq
to relate the cohomology classes to the ones on the ambient fourfold $A=\IP^4_{11114}$. Now, using the algorithm of \cite{Distler:1996tj, Blumenhagen:1997vt} to compute cohomology classes of line bundles over toric varieties, we find that the only non-vanishing cohomology classes are $H^4\left(A, {\cal O}_A(-n)\right)$ and $H^4\left(A, {\cal O}_A(-n-8)\right)$, which are generated by the following local sections in \v{C}ech cohomology:
\beq
  \bal
    H^4\left(A, {\cal O}_A(-n)\right)  &:\ \frac{1}{u_1\, u_2\, u_3\, u_4\, \xi\, P_{n-8}(\vec u,\xi)}\\
    H^4\left(A, {\cal O}_A(-n-8)\right)&:\ \frac{1}{u_1\, u_2\, u_3\, u_4\,\xi\, P_{n}(\vec u,\xi)}\; .  
  \eal
\eeq
where $P_n$ denotes a polynomial of degree $n$ in the homogeneous coordinates. The number $p_n$ of such polynomials can be written as
\beq
 p_n= \sum_{k=0}^{[n/4]} \binom{n-4k+3}{3}
\eeq
so that the Koszul sequences finally give
\beq 
  h^*(E_n,{\cO}_{E_n})=(1,0,p_n-p_{n-8}-1) \; .
\eeq
One can show that this is consistent with the holomorphic Euler characteristic in eq.~\eqref{eulerchar}. Therefore none of the divisors $E_n$ admit any non-trivial Wilson lines but generically do have transverse deformations.

Since we are considering $O(1)$-instantons, it is not sufficient to compute $h^{0,i}$ but instead we require the $\sigma$-invariant and anti-invariant parts of these Hodge numbers, $h^{0,i}_\pm(E_n)$. Since we know already that $h^1(E_n,{\cO}_{E_n})=0$, these individual numbers can simply be computed by the Lefschetz index theorem from appendix \ref{app:index}. We have already computed $\chi(E_n,{\cal O}_{E_n})$. For the index $\chi^\sigma (E_n,{\cal O}_{E_n})$ one obtains
\beq\label{eulerequiva}
  \chi^{\sigma} (E_n,{\cal O}_{E_n})= -2\cdot\frac{1}{4}\,n^2\cdot 4 = -2\,n^2\; .
\eeq
Serre duality tells us that $h^{0,2}$ equals the number of sections of the normal bundle, so we can independently count these and deduce the Hodge numbers.\footnote{Note that Serre duality relates $h^{0,2}(E, \mathcal{O}) = h^0(E, \mathrm{N}_E)$ via contraction of the section of the normal bundle with the holomorphic three-form $\Omega$. However, since the latter is odd under the orientifold action, this means that $\sigma$-\emph{invariant} sections of the normal bundle are counted by $h^{0,2}_-$, and $\sigma$-\emph{odd} sections are counted by $h^{0,2}_+$. For this simple model, the reader can easily check this explicitly for low $n$.} Hence, we automatically deduce the following:
\beq
  \bal
    h^{0,2}_+(E_n) &= \frac{1}{6}\,(n^3+11\,n)-n^2-1 \\
    h^{0,2}_-(E_n) &= \frac{1}{6}\,(n^3+11\,n)+n^2\;,
  \eal
\eeq
which we can also summarize as
\beq\label{superfrak}
  \fh(E_n)=\left(1,0,{\textstyle \binom{n-1}{3}},{\textstyle \binom{n+3}{3}}-1\right)\; .
\eeq

\subsection[\texorpdfstring{$U(1)$}{U(1)} E3-brane instantons]{\boldmath$U(1)$ E3-brane instantons}

Now let us choose $n=2m$ for $m\ge 4$  and consider a brane/image-brane pair of Euclidean E3-branes
\beq
   E3:\ (\eta_{m} + \xi\, \psi_{m-4})\; , \qquad\quad
  E3':\ (\eta_{m} - \xi\, \psi_{m-4})=0\; .
\eeq     
For this system to have the same charges with respect to the R-R $C_4$ and $C_0$ form as the recombined $O(1)$ instanton, the world-volumes of the two branes must also carry non-trivial line bundles
\beq
  { E3}:\  L ={\cal O}\left(-{\textstyle\frac{m}{2}}\right), \qquad\quad 
  { E3'}:\ L'={\cal O}\left({\textstyle\frac{m}{2}}\right)\; .
\eeq 
Note that this is also the right gauge flux to cancel the Freed-Witten anomaly in the case that $m$ is odd.

Since the two instantons are exchanged under the orientifold projection, one gets the zero modes
\beq
  h^{0}_\pm (E_m,{\cal O}_{E_m})=1, \qquad
  h^{2}_\pm (E_m,{\cal O}_{E_m})=\frac{1}{3}\,(m^3+11\,m)-1
\eeq
and $h^{1}_\pm (E_m,{\cal O}_{E_m})=0$. The extra universal zero mode $h^{0}_- (E_m,{\cal O}_{E_m})=1$ indicates that this is a $U(1)$ instanton. We can also write these contributions as
\beq
  \fh(E3-E3)=\left(1,1,{\textstyle\frac{1}{3}\,(m^3+11\,m)-1},{\textstyle
      \frac{1}{3}\,(m^3+11\,m)-1}\right)\; .
\eeq

In this case one expects additional zero modes on the intersection of the E3-brane and its orientifold image E3'. The E3-instanton and its orientifold image intersect over the curve $C=E3 \cap E3'$ with $K_C={\cal O}(2m)$. Analogous to the previously discussed recombination of the D7$-$D7' system, by definition the homology class of this curve is invariant under $\sigma$ so that we can split the cohomology classes in $\sigma$ invariant and anti-invariant pieces. Let us compute\footnote{Since $C$ is not smooth, one has again to apply the formalism of \cite{Blumenhagen:2006wj}, which gives the same result.}
\beq\label{recomzeroprop}
  H^i_\pm(C, L^2\otimes K_C^{\frac{1}{2}})=   H^i_\pm (C, {\cal O}_C)\; , \quad i=0,1\; .
\eeq
The holomorphic Euler characteristic of the curve $C$ can readily be computed as
\beq\label{indexlala}
  \chi(C, {\cal O}_C)=-2\, m^3\; ,
\eeq 
from which we conclude with $h^0(C,{\cal O}_C)=1$ that $h^1(C,{\cal O}_C)=2m^3+1$.

The chiral index for the invariant and anti-invariant cohomologies can be computed via eq.~\eqref{chiralantiindex}
\beq
  h^0_-(C,{\cal O}_C)-h^1_-(C,{\cal O}_C) = -m^2\, (m-4)\; ,
\eeq
with \eqref{indexlala} leading to 
\beq
  h^0_+(C,{\cal O}_C)-h^1_+(C,{\cal O}_C) = -m^2\, (m+4)\; .
\eeq
With $h^{0}_+ (C,{\cal O}_{C})=1$  we can conclude
\beq
  \bal
    &h^{0}_+ (C,{\cal O}_{C})=1  \qquad\qquad\qquad\phantom{aaa}   h^{0}_- (C,{\cal O}_{C})=0\\
    &h^{1}_+ (C,{\cal O}_{C})=m^3+4m^2+1 \qquad   
     h^{1}_- (C,{\cal O}_{C})=m^3-4m^2\; .    
  \eal 
\eeq
To see what is happening, let us put these numbers in another $\fh$ four-vector       
\beq
  \bal
    \fh(E3-E3')&=\Bigl(0\, ,\,  h^{0}_-(C,{\cal O}_C)\, , \, 
    h^{0}_+(C,{\cal O}_C) + h^{1}_-(C,{\cal O}_C)\, , \,
    h^{1}_+(C,{\cal O}_C) \Bigr)\\
    &= \left(0,0,m^3-4m^2+1, m^3+4m +1 \right)\; .
  \eal
\eeq
Now, recombining this E3$-$E3' system, one pair of modes pairs up. Introducing $\fh({\rm recomb})=(0,1,1,0)$, we obtain for the total number of zero modes
\beq
  \bal
    \fh(E3-E3)+{}&\fh(E3-E3')-\fh({\rm recomb})=\\
    &{}=\left(1,0,{\textstyle \frac{4}{3} m^3 -4m^2 +\frac{11}{3}m-1 },
    {\textstyle \frac{4}{3} m^3 +4m^2 +\frac{11}{3}m }\right) \\
    &{}=\left(1,0,{\textstyle \binom{2m-1}{3}},{\textstyle
    \binom{2m+3}{3}}-1\right) \\
    &{}= \fh(E_{2m})\; .
  \eal
\eeq
Thus the dimensions of instanton moduli spaces of the single $O(1)$ instanton and the recombined $U(1)$ instanton pair perfectly match.

\subsection{F-theory uplift}

Next we consider the F-theory uplift to a Calabi-Yau fourfold $\cZ$ of this simple orientifold model. It is given by the Weierstra\ss\ fibration over the threefold base $\IP^3$. By adding the elliptic fibration in the form of a $\IP_{231}^2[6]$-bundle, the Calabi-Yau fourfold $\cZ$ can be described as a degree-6 hypersurface in a toric ambient fivefold ${\cal A}$. The toric data are presented in Table \ref{tbl:PthreeUplift}.

\begin{table}[ht]
  \begin{center}
    \begin{tabular}{r@{\,$=$\,(\,}r@{,\;\;}r@{,\;\;}r@{,\;\;}r@{,\;\;}r@{\,)\;\;}|c|cc|c} 
      \multicolumn{6}{c|}{vertices of the} & coords & \multicolumn{2}{c|}{GLSM charges} & {divisor class}${}^\big.$ \\
      \multicolumn{6}{c|}{polyhedron / fan}      &        & $Q^1$   &   $Q^2$  &  \\ 
      \hline\hline
      $\rho_1$ &   0  &   0  &   0  &   1  &   0  &  $x$  & 2 &   0  & $2\sigma+8H$\uspc \\
      $\rho_2$ &   0  &   0  &   0  &   0  &   1  &  $y$  & 3 &   0  & $3\sigma+12H$ \\
      $\rho_3$ &   0  &   0  &   0  & $-2$ & $-3$ &  $z$  & 1 & $-4$ & $\sigma$ \\
      $\rho_4$ & $-1$ & $-1$ & $-1$ & $-8$ & $-12$& $u_1$ & 0 &   1  & $H$\uspc \\
      $\rho_5$ &   1  &   0  &   0  &   0  &   0  & $u_2$ & 0 &   1  & $H$  \\
      $\rho_6$ &   0  &   1  &   0  &   0  &   0  & $u_3$ & 0 &   1  & $H$  \\
      $\rho_7$ &   0  &   0  &   1  &   0  &   0  & $u_4$ & 0 &   1  & $H$\lspc  \\ 
      \hline
      \multicolumn{6}{c|}{conditions:}  &                 & 6 &   0  & \uspc 
    \end{tabular}
  \end{center}
  \caption{\small Toric data for the F-theory uplift fourfold over $\IP^3$.}
  \label{tbl:PthreeUplift}
\end{table}

Using standard methods of toric geometry, we can compute various topological quantities. First, the number of K\"ahler moduli is $h^{1,1} = 2$. Choosing as a basis of the Poincar\'e dual $H_{6}(\cZ,\IZ)$ the divisors $\sigma=\{z=0\}$ (the embedding of the $\IP^3$ basis) and $H=\{u_1=0\}$ (the pull-back $\pi^{-1}(\IP^2)$), the intersection form can be computed as
\beq
  I_{\cZ} = -64 \sigma^4 + 16 \sigma^3 H - 4 \sigma^2 H^2 + \sigma H^3.
\eeq
From this one determines $\chi(\cZ) = 23328$. Moreover, there are $h^{3,1} = 3874$ complex structure deformations as well as $h^{2,1} = 0$ and $h^{2,2}=15564$.

Now, we consider  the divisors  $\cM_n=n H$, which are nothing else than $\pi^{-1}(\cE_n)$, i.e.~the pull-back of the divisors $\cE_n$ in the base $\IP^3$, where $\cE_n = E_n/\sigma$. The holomorphic Euler characteristic of any threefold divisor $\cM_n$ can be readily computed by
\beq\label{eulercharff}
  \bal
    \chi(\cM_n,{\cO}_{\cM_n})&= \sum_{i=0}^3 (-1)^i h^i(\cM_n,{\cal O}_{\cM_n})\\
    &= \frac{1}{24}\int_{\cM_n}  c_1(\cM_n)\, c_2(\cM_n) \\
    &= -2 n^2 \; ,
  \eal
\eeq
which looks completely different from the Euler characteristic of $E$ in eq.~\eqref{eulerchar} but apparently equals precisely the $\mathbb Z_2$ Euler characteristic $\chi^\sigma(E,{\cal O}_{E})$ in \eqref{eulerequiva}.

To see what is going on, we have to compute not only the index but the individual cohomology classes $H^i(\cM_n,{\cal O}_{\cM_n})$. For this purpose we use the Koszul sequence
\beq
  0\fto {\cal O}_{\cZ}(0,-n)\injto {\cal O}_{\cZ} \surjto {\cal O}_{\cM_n} \fto 0
\eeq
and the associated long exact sequence in cohomology to relate $H^i(\cM_n,{\cal O}_{\cM_n})$ to line-bundle cohomologies on the Calabi-Yau fourfold $\cZ$. For the cohomology of the structure sheaf on $\cZ$ one simply finds $h^i(\cZ,{\cal O}_{\cZ})=h^{i,0}(\cZ)=(1,0,0,0,1)$. To compute the cohomologies of ${\cal O}_{\cZ}(0,-n)$, we employ a second Koszul sequence
\beq\label{sequencea}
  0  \fto {\cal O}_{\cal A}(-6,-n) \injto {\cal O}_{\cal A}(0,-n) \surjto {\cal O}_{\cal \cZ}(0,-n)  \fto 0
\eeq
to relate them to the cohomology classes of line bundles on the toric ambient fivefold ${\cal A}$, which is described by the toric data in Table \ref{tbl:PthreeUplift}. Using the combinatorial algorithm from \cite{Distler:1996tj, Blumenhagen:1997vt}, one finds that the cohomology of the line bundle ${\cal O}(-6,-n)$ receives contributions only to the maximal $H^5({\cal A}, {\cal O}(-6,-n))$. The elements arise from local sections
\beq
  H^5\big({\cal A}, {\cal O}(-6,-n)\big):\ \frac{1}{y x z u_1 u_2 u_3 u_4\, P_n(\vec u)}
\eeq
in \v{C}ech cohomology. Similarly, one finds that for the second line bundle in \eqref{sequencea} only $H^3({\cal A}, {\cal O}(0,-n))\ne 0$, where the elements consists of local sections 
\beq
  H^3\big({\cal A}, {\cal O}(0,-n)\big):\ \frac{1}{u_1 u_2 u_3 u_4 \, P_{n-4}(\vec u)}\; .
\eeq
Therefore, we obtain for the dimensions of the two line bundle cohomologies
\beq
  \bal
    h^*\big({\cal A},{\cal O}(-6,-n)\big)&=\left(0,0,0,0,0,{\textstyle \binom{n+3}{3}}\right), \\
    h^*\big({\cal A},{\cal O}(0,-n)\big)&=\left(0,0,0,{\textstyle \binom{n-1}{3}},0,0\right)\;.
  \eal
\eeq
Now, running back through the long exact sequences in cohomology for the two Koszul sequences, one uniquely finds
\beq\label{ftheoryresa}
  h^*(\cM_n,{\cal O}_{\cM_n})=\left( 1,0, {\textstyle \binom{n-1}{3}},{\textstyle \binom{n+3}{3}}-1\right) \; ,
\eeq       
which is consistent with the holomorphic Euler characteristic \eqref{eulercharff}.

We would like to compare this data with the corresponding Type IIB E3-instanton zero modes. It is obvious that the F-theory zero modes of the divisor ${\cal M}_n$ precisely match with $\fh(E_n)$ in eq.~\eqref{superfrak}. Therefore, we have the relations shown in Table \ref{tabledeformc} between the equivariant cohomology of the $O(1)$ Type IIB divisor $E$ and its F-theory pull-back ${\cal M}$.
\begin{table}[ht]
  \centering
  \begin{tabular}{c|c|c|c}
    zero modes                    &  statistics  & Type IIB            & F-theory      \\ 
    \hline\hline
    $(X_\mu,\theta_\alpha)$       & (bose, fermi) & $H_+^{0,0}(E)$ & $H^{0,0}(\cM)$\uspc\lspc \\  
    \hline
    $\ov\tau_{\dot\alpha}$        & fermi        & $H_-^{0,0}(E)$\uspc & \multirow{2}{*}{$H^{1,0}(\cM)$} \\
    $\gamma_{\alpha}$             & fermi        & $H_+^{1,0}(E)$\lspc &      \\
    \hline
    $(w, \ov\gamma_{\dot\alpha})$ & (bose, fermi) & $H_-^{1,0}(E)$\uspc & \multirow{2}{*}{$H^{2,0}(\cM)$} \\
    $\chi_{\alpha}$               & fermi & $H_+^{2,0}(E)$\lspc & \\
    \hline
    $(c , \ov\chi_{\dot\alpha} )$ & (bose, fermi)        & $H_-^{2,0}(E)$ & $H^{3,0}(\cM)$\uspc
  \end{tabular}
  \caption{\small Type IIB and F-theory zero modes for $O(1)$ instanton.}
  \label{tabledeformc}
\end{table} 
How can we understand this uplift of the cohomologies? The answer is roughly the following: The presence of the O7-plane induces a monodromy on the homology of the elliptic fiber of the form $(A, B) \mapsto (-A, -B)$, for $A, B \in H^2(T^2, \mathbb{Z})$. This means that an $(i,0)$-form of the ``upstairs'' D3-brane in $X$ must be $\sigma$-odd in order to combine with the $(1,0)$-form of the fiber into an $\sigma$-even $(i+1,0)$-form. Only such a form will survive the geometric quotient that gives us the F-theory fourfold $\smash{\cZ \stackrel{\pi}{\surjto} \cB=X/\sigma}$. The uplift of the cohomologies is summarized in figure \ref{fig:cohomologyflow}

\begin{figure}[h] 
\begin{center}
\begin{tikzpicture}[scale=1, node distance = 2cm, auto, inner sep=.1mm, text width=1.5cm, text centered]
\node (02m) at (0,.25) {$h^{0,2}_-(E3)$} ;
\node (02p) at (0,-.3) {$h^{0,2}_+(E3)$};
\node (01m) at (0,-1.25) {$h^{0,1}_-(E3)$} ;
\node (01p) at (0,-1.8) {$h^{0,1}_+(E3)$};
\node (00m) at (0,-2.75) {$h^{0,0}_-(E3)$} ;
\node (00p) at (0,-3.3) {$h^{0,0}_+(E3)$};
\node (03) at (4,1.3) {$h^{0,3}(M5)$};
\node (02) at (4,-.2) {$h^{0,2}(M5)$};
\node (01) at (4,-1.7) {$h^{0,1}(M5)$};
\node (00) at (4,-3.2) {$h^{0,0}(M5)$};
\path [line] (00p) -- (00);
\path [line] (00m) -- (01);
\path [line] (01p) -- (01);
\path [line] (01m) -- (02);
\path [line] (02p) -- (02);
\path [line] (02m) -- (03);

\end{tikzpicture}
\end{center}
\caption{The even cohomologies of the E3 travel horizontally, whereas the odd ones are raised to forms of the M5 of one degree higher .}
\label{fig:cohomologyflow}
\end{figure}

We conclude that this non-trivial computation convincingly shows the equivalence of E3-brane instantons in Type IIB orientifolds and vertical M5-brane instantons in F-theory. In particular this implies that in F-theory $h^{i,0}(M)$ does only count the universal and the deformation zero modes, which for E3-brane instantons are described by open strings starting and ending on the $O(1)$ instantonic brane. From orientifolds we know that there can be additional matter zero modes, arising from open strings from the E3-brane instanton to space-time filling D7-branes. In F-theory these must come from intersections of the divisor $\cM$ with the components of the discriminant locus, over which the elliptic fiber degenerates. We will discuss this important point in more detail in section \ref{sec_matterzero}.

Let us finish this section with some comments on the uplift of $U(1)$ E3-instantons. In the previous subsection we have discussed the splitting of the $O(1)$ instanton wrapping $E_{2m}$ into a pair of E3$-$E3' branes each wrapping $E_m$. This is a $U(1)$ instanton, as seen, carrying additional zero modes in the E3$-$E3' sector which can recombine the brane/image-brane. In F-theory this brane/image-brane configuration corresponds to the restricted divisor
\beq
  \eta_m^2 - h\, \psi_{m-4}^2 =0
\eeq
which is singular at $\eta=\psi=0$. Even resolving the singularity, we did not find the $\ov\tau_{\dot\alpha}$ zero modes, so that in F-theory only the recombined $O(1)$ instanton is present. We would like to view this as a non-perturbative effect, which smooths out this singularity in the perturbative $E_{2m}$ moduli space.

Then the question arises whether there can exist $U(1)$ Type IIB instantons, which do not non-perturbatively recombine in the F-theory uplift. From the discussion so far, for this to be the case we need an E3-brane wrapping a four-cycle which does not intersect the orientifold plane. Moreover, from Table \ref{tabledeformc} we expect that in this case the universal Type IIB $\ov\tau_{\dot\alpha}$ zero mode appears as a non-trivial element in $H^{1,0}(\cM)$ in the F-theory uplift. It is the aim of the next section to construct such an example.

\subsection[Non-recombining \texorpdfstring{$U(1)$}{U(1)} instantons in F-theory]{Non-recombining \boldmath$U(1)$ instantons in F-theory}

In this subsection we will provide an example of a Type IIB orientifold and its F-theory uplift which contains a $U(1)$ instanton whose universal zero modes $\ov\tau_{\dot\alpha}$ are not non-perturbatively lifted in F-theory, in order to find out to what mode of the M5-brane they correspond. The considerably simple Calabi-Yau geometries considered so far are not sufficient for this purpose so that we need to introduce a second example of a Calabi-Yau three- and fourfold.\footnote{The reader not so familiar with toric geometry might directly jump to the next section, where only the example from the previous sections is needed.}

Our aim is to define a geometry containing an E3$-$E3' brane pair, which do not intersect each other. This means in particular that the E3-brane does not intersect the O7-plane. Then the D7-brane tadpole cancellation condition implies that, for only a single $SO(1)$ D7-brane present, the E3-brane does not intersect the D7-brane either. Therefore, in the F-theory uplift there is no degeneration of the elliptic fiber on the base of the M5-brane $T^2\injto \cM\surjto \cE$. This means that its topology is $\cM=T^2\times \cE$ and hence the holomorphic Euler characteristic vanishes, as every element $\omega^i\in H^{i,0}(\cE)$ gives rise to both 
\beq
  \omega^i\in    H^{i,0}(\cM) \qquad {\rm and}\quad \omega^i\wedge \omega_{T^2}\in    H^{i+1,0}(\cM)
\eeq
for $\omega_{T^2}\in H^{1,0}(T^2)$. Type IIB orientifolds with involutions exchanging two divisors have been considered in \cite{Blumenhagen:2009up,Collinucci:2009uh}. A still very simple example of this type is provided by starting with the quintic $\IP^4[5]$ and performing two del Pezzo transitions. Then one can define an involution $\sigma$ which exchanges these two del Pezzo surfaces. Let us define this geometry in more detail, additional information can be found in \cite{Blumenhagen:2009up,Collinucci:2009uh}.

\subsubsection*{The upstairs Type IIB orientifold}
The toric data of our upstairs Calabi-Yau threefold is summarized in Table \ref{tbl:dPtransCY3}. For such toric varieties there arise various phases (triangulations), where the geometric ones --- usually those with the maximal number of cones --- are related via flop transitions. In our case the toric ambient space has just a single such phase $\Sigma_{12}$ with 12 cones. This is the geometry discussed in detail in \cite{Blumenhagen:2008zz}, where one finds two $dP_7$ surfaces intersecting over a $\IP^1$. Since these two $dP_7$s are intersecting each other, they are not suitable for F-theory $U(1)$ divisors of the non-recombining type we are currently looking for. In fact, wrapping an E3-brane on both divisors lifts to a single, smooth M5-brane that satisfies Witten's strong criterion $h^{0,i}(\mathcal{M})={(1, 0 ,0 ,0)}$ !

\begin{table}[ht]
  \begin{center}
    \begin{tabular}{r@{\,$=$\,(\,}r@{,\;\;}r@{,\;\;}r@{,\;\;}r@{\,)\;\;}|c|ccc|c} 
      \multicolumn{5}{c|}{vertices of the} & coords & \multicolumn{3}{c|}{GLSM charges} & {divisor class}${}^\big.$ \\
      \multicolumn{5}{c|}{polyhedron / fan} & & $Q^1$ & $Q^2$ & $Q^3$ & \\ 
      \hline\hline
      $\nu_1$ & $-1$ & $-1$ & $-1$ & $-1$ & $u_1$ & 1 & 0 & 0 & $H$\uspc \\
      $\nu_2$ &   1  &   0  &   0  &   0  & $u_2$ & 1 & 0 & 0 & $H$ \\
      $\nu_3$ &   0  &   1  &   0  &   0  & $u_3$ & 1 & 0 & 0 & $H$ \\
      $\nu_4$ &   0  &   0  &   1  &   0  & $v_1$ & 1 & 0 & 1 & $H+B$ \\
      $\nu_5$ &   0  &   0  &   0  &   1  & $v_2$ & 1 & 1 & 0 & $H+A$ \\
      $\nu_6$ &   0  &   0  &   0  & $-1$ & $w_1$ & 0 & 1 & 0 & $A$ \\
      $\nu_7$ &   0  &   0  &  $-1$  & 0  & $w_2$ & 0 & 0 & 1 & $B$\lspc \\ 
      \hline
      \multicolumn{5}{c|}{conditions:}    &       & 5 & 2 & 2 & \uspc \\
    \end{tabular}
  \end{center}
  \caption{\small Toric data for the double dP-transition of the quintic threefold $X=\IP^4[5]$.}
  \label{tbl:dPtransCY3}
\end{table}

As we now argue, there exists a ``non-maximal''\footnote{By ``maximal'' we mean the triangulation that creates the highest amount of cones in the fan. For this polytope, this would be the $\Sigma_{12}$ triangulation with 12 cones.} triangulation (phase) $\Sigma_{11}$ of the ambient space with 11 cones, which, when restricted to the Calabi-Yau hypersurface, is nevertheless related by a flop transition to the ``maximal'' phase. It is very convenient to  describe this transition in two-dimensional Gauged Linear Sigma Model (GLSM) language. Starting with the GLSM charges listed in Table \ref{tbl:dPtransCY3}, the D-term constraints for the three two-dimensional $U(1)$ are:
\beq
  \bal
    |u_1|^2+|u_2|^2+|u_3|^2-|w_1|^2-|w_2|^2 &= \xi_1\,,\\
    |v_2|^2+|w_1|^2 &= \xi_2\,,\\
    |v_1|^2+|w_2|^2 &= \xi_3\,.
  \eal
\eeq
We impose $\xi_2, \xi_3>0$, $\xi_1+\xi_2>0$, and $\xi_1+\xi_2>0$. The ``maximal'' triangulation/fan $\Sigma_{12}$ corresponds to $\xi_1>0$. We will now consider the transition from $\xi_1>0$ to $\xi_1<0$. For $\xi_1>0$ the set $u_1=u_2=u_3$ is disallowed so that the SR-ideal must contain $u_1 u_2 u_3$. For $\xi_1\!\!\searrow\! 0$ in the ambient toric variety, the surface $w_1=w_2=0$ shrinks to zero size. This is a $\IP^2$. Now, for $\xi_1<0$ the set $w_1=w_2=0$ is excluded, i.e.~the SR-ideal  contains $w_1 w_2$. For $\xi_1\!\!\nearrow\! 0$, the curve $u_1=u_2=u_3=0$, which is a $\IP^1$, shrinks the zero size. We conclude that in the ambient space at the boundary of the K\"ahler cone we have a transition from a $\IP^2$ to a $\IP^1$. This is not a flop transition, consistent with the fact the two related triangulations have different number of cones. It can be checked that this phase is geometric and K\"ahler.

However, we still have to restrict this to the Calabi-Yau hypersurface constraint. Here we find that
\beq
  \xi_1>0:\ \IP^2\cap [5H+2A+2B]=\IP^1, \qquad
  \xi_1<0:\ \IP^1\subset  [5H+2A+2B]\; ,
\eeq
so that  we have a $\IP^1$ in both phases and they are related by a flop transition. We conclude that even though the phase $\Sigma_{11}$ is not maximal on the ambient space, it gives a well defined geometric phase on the Calabi-Yau hypersurface, which is related to $\Sigma_{12}$ by a flop transition.

As mentioned, in \cite{Blumenhagen:2008zz} the phase $\Sigma_{12}$ has been discussed in detail and it was found that the divisors $A$ and $B$ are two non-generic intersecting $dP_7$ surfaces. As we have seen above, in $\Sigma_{11}$ the intersection curve $\IP^1$ is flopped, so that $A$ and $B$ are two non-intersecting $dP_6$ surfaces. Let us further discuss this latter phase.

In the basis $H \ce D_{u_1}$, $A \ce D_{w_1}$ and $B \ce D_{w_2}$ the non-zero triple intersection numbers are 
\beq
  I_{\QdPsix} = -H^3+3\big( A^3 + B^3 - H(A^2+B^2) + H^2(A+B) \big),
\eeq
and the Stanley-Reisner ideal which is modded out from the intersection ring is
\beq
  \mathrm{SR}(\QdPsix) = \langle v_1 w_2, \;  v_2 w_1, \;  w_1w_2, \;  u_1 u_2 u_3 v_1, \; u_1 u_2 u_3 v_2 \rangle.
\eeq
Consistently, we find that in the phase $\QdPsix$ both divisors $D_{w_1}$ and $D_{w_2}$ have Euler characteristic $\chi(D_{w_i}) = 9$. 

Now we want the two non-intersecting $dP_6$ surfaces to support our pair of E3$-$E3' instantons. Therefore we need to impose an orientifold projection which exchanges them. The globally well defined involution doing this is
\beq\label{involution_dP6}
  \sigma: \begin{cases}  v_1\leftrightarrow v_2, \\
                         w_1\leftrightarrow w_2. \end{cases} 
\eeq 
Using the projective equivalences we find that the O7-plane corresponds to the fixed point locus defined by
\beq
  v_1 w_1 - v_2 w_2 = 0,
\eeq
which describes a surface in the class $[H+A+B] \in H_4(X,\IZ)$, for which we compute $\chi(O7)=55$ for $\QdPsix$. Furthermore, the isolated fix-point $(0,0,0,1,-1,1,1)\in X$ corresponds to the location of a single $O3$-plane.

\subsubsection*{F-theory uplift}
As before we are looking for a projection mapping from the Calabi-Yau threefold $X$ to the quotient geometry $\cB = X / \sigma$, which is 2-to-1 away from the O3/O7-plane and 1-to-1 on it. Such a mapping can be defined via
\beq\label{eq:QdPsixProjMapping}
  (u_1,u_2,u_3,v_1,v_2,w_1,w_2) \mapsto (u_1,u_2,u_3,\underbrace{v_1v_2}_v, \underbrace{w_1w_2}_w,\underbrace{v_1w_1+v_2w_2}_h)
\eeq
with the new ``downstairs'' coordinates defined as indicated. In this quotient geometry the location of the O7-plane corresponds to the divisor $h=0$ and the $O3$-plane is found at $(0,0,0,-1,1,0)\in\cB$. The toric data of this base manifold and the elliptic $\IP^2_{2,3,1}[6]$-fibration over it is described by the toric data in Table \ref{tbl:dPtransUplift}. The Calabi-Yau fourfold $\cZ$ is then given as the complete intersection of two polynomials
\beq
  \{ P^{(6,0,0)} = 0 \} \cap \{ W^{(0,5,2)} = 0 \}
\eeq
with the respective degrees indicated.

\begin{table}[ht]
  \centering
  \begin{tabular}{r@{\,$=$\,(\,}r@{,\;\;}r@{,\;\;}r@{,\;\;}r@{,\;\;}r@{,\;\;}r@{\,)\;\;}|c|ccc|c} 
    \multicolumn{7}{c|}{vertices of the} & coords & \multicolumn{3}{c|}{GLSM charges} & {divisor class} \\
    \multicolumn{7}{c|}{polyhedron / fan} & & $Q^1$ & $Q^2$ & $Q^3$ & \\ 
    \hline\hline
    $\mu_1$ &   0  &   0  &   0  &   0  &   1  &   0  &  $x$  & 2 &  0  &  0  & $2\sigma+2P+2A$\uspc \\
    $\mu_2$ &   0  &   0  &   0  &   0  &   0  &   1  &  $y$  & 3 &  0  &  0  & $3\sigma+3P+3A$ \\
    $\mu_3$ &   0  &   0  &   0  &   0  & $-2$ & $-3$ &  $z$  & 1 &$-1$ &$-1$ & $\sigma$ \\
    $\mu_4$ & $-1$ & $-1$ & $-2$ & $-1$ & $-2$ & $-3$ & $u_1$ & 0 &  1  &  0  & $P$\uspc \\
    $\mu_5$ &   1  &   0  &   0  &   0  &   0  &   0  & $u_2$ & 0 &  1  &  0  & $P$  \\
    $\mu_6$ &   0  &   1  &   0  &   0  &   0  &   0  & $u_3$ & 0 &  1  &  0  & $P$  \\
    $\mu_7$ &   0  &   0  &   1  &   0  &   0  &   0  &  $v$  & 0 &  2  &  1  & $2P+A$  \\
    $\mu_8$ &   0  &   0  &   0  &   1  &   0  &   0  &  $h$  & 0 &  1  &  1  & $P+A$  \\
    $\mu_9$ &   0  &   0  & $-1$ & $-1$ & $-2$ & $-3$ &  $w$  & 0 &  0  &  1  & $A$\lspc  \\ 
    \hline
    \multicolumn{7}{c|}{conditions:}  &                       & 6 & 0 & 0 & \uspc \\
    \multicolumn{7}{c|}{ }            &                       & 0 & 5 & 2 & 
  \end{tabular}
  \caption{\small F-theory uplift fourfold of the exchange involution orientifold of $\QdPsix$.}
  \label{tbl:dPtransUplift}
\end{table}

From those vertices we can construct three consistent fans $\tilde\Sigma_{28}$, $\tilde\Sigma_{27}$ and $\tilde\Sigma_{24}$ of the sixfold ambient space of the uplift fourfold. The first two correspond to different uplift variants of $\Sigma_{12}$ with $dP_7$-surfaces, whereas $\tilde\Sigma_{24}$ uplifts the Calabi-Yau threefold $\QdPsix$ associated to the fan $\Sigma_{11}$ with a $dP_6$-surface.

We focus on the last case of $\tilde\Sigma_{24}$, which will be denoted as $\UpQdPsix$. The base $\cB$ is again embedded as the divisor $z=0$, which defines a threefold in the class $[\sigma]\in H_6(\cZ;\IZ)$. The non-vanishing intersection numbers of the Calabi-Yau fourfold are
\beq
  \textstyle I_{\UpQdPsix} = -\frac{5}{2}\sigma^4 + \frac{5}{2}\sigma^3 P - \frac{5}{2}\sigma^2 P^2 - \sigma\left(\frac{1}{2} P^3 + 3 P^2 A - 3 P A^2 + 3 A^3 \right)
\eeq
where the half-integers hint at the presence of an unresolved $\IZ_2$-singularity, which is related to the O3-plane identified in the Type IIB geometry. The Stanley-Reisner ideal
\beq
  \mathrm{SR}(\UpQdPsix) = \langle h w, \;  x y z, \;  u_1 u_2 u_3 v \rangle
\eeq
then defines the intersection ring.

\begin{table}[ht]
  \centering
  \begin{tabular}{r||ccccccc}
    coords: & $u_1$ & $u_2$ & $u_3$ & $v$ & $x$ & $y$& $z$ \\ \hline
    $Q_1'$: & 0 & 0 & 0 & 0 & 2 & 3 & 1\uspc \\
    $Q_2'$: & 1 & 1 & 1 & 1 & 0 & 0 & 0 
  \end{tabular}
  \caption{GLSM charges of gauge-fixed, simplified ambient space.}
  \label{tab_simplified}
\end{table}

We now want to study the properties of the divisor $w=0$, whose pre-image under the orientifold projection mapping \eqref{eq:QdPsixProjMapping} is the disjoint pair $\{ w_1=0 \} \cup \{ w_2=0 \}$ of $dP_6$-surfaces. Given the Stanley-Reisner ideal for the ambient sixfold, we can gauge-fix $h$ to one, and divide $v$ by $h$. The resulting manifold is then the ambient space described by the GLSM charges in Table \ref{tab_simplified} with Stanley-Reisner ideal
\beq
  \mathrm{SR} = \langle u_1 u_2 u_3 v, \; x y z \rangle\,,
\eeq
which describes the fivefold $\IP^3\times\IP^2_{231}$. The threefold in question is then given by the intersection of two equations 
\beq
  \{ P^{(6,0)}=0 \} \cap \{ W^{(0,3)}=0 \}\,.
\eeq
We easily recognize this as the product of a cubic in $\IP^3$, and an elliptic curve. In other words
\beq
  D_w \cong dP_6 \times T^2
\eeq
with Hodge numbers:
\beq
  h^{0,i} = (1,1,0,0)\,.
\eeq
As expected, since the two $dP_6$ surfaces on the upstairs Calabi-Yau threefold were non-intersecting, the F-theory uplift gives a $dP_6$ surface $dP_6\subset {\cal B}$, over which the elliptic fiber does not degenerate. Therefore, both the fermionic zero modes $\theta_{\alpha}$ counted by $h^{0,0}_+ (dP_6)=1$ and the modes $\ov\tau_{\dot\alpha}$ counted by $h^{0,0}_- (dP_6)$ survive in the F-theory uplift. This M5-brane instanton can therefore be considered as a $U(1)$ instanton, which proves the existence of non-recombining/``genuine'' $U(1)$ instantons in the non-perturbative F-theory framework.

This examples teaches us that the $\ov\tau_{\dot\alpha}$ zero-modes of the E3-brane contribute to $h^{0,1}(\mathcal{M})$.

\section{Charged zero modes}\label{sec_matterzero}

In the previous sections we have considered the E3-instanton zero modes arising from open strings starting and ending on the E3-brane (respectively for $U(1)$ instantons on E3 and its orientifold image E3'). Via the Sen-limit of the corresponding F-theory uplift we have managed to relate the Type IIB instanton zero modes to the zero modes of appropriate M5-brane instantons in F-theory. By this quite explicit procedure we were able to confirm the results of \cite{Witten:1996bn}, namely that these zero modes are counted by the cohomology classes $H^i(\cM,{\cal O}_\cM)$, $i\in\{0,\ldots,3\}$ on the six-dimensional divisor wrapped by the Euclidean M5-brane.

We know from Type IIB orientifolds that these universal and deformation zero modes are only half of the story. Indeed from the intersection of the E3-branes with the space-time filling D7-branes, there generically appear fermionic so-called matter zero modes. These are counted by the cohomology classes $\smash{H^i(C,L\otimes K_C^{1/2})}$ with $i\in\{0,1\}$, where $C$ is the intersection curve of the E3-divisor and the internal D7-divisor. Moreover, $L$ denotes an additional line bundle present on the D7-brane. Note that these additional fermionic zero modes cannot be neglected, as they eventually determine to which effective operator the E3-instanton can contribute.

Clearly, these fermionic matter zero modes must have a correspondence for M5-brane instantons in F-theory. Noting that in F-theory the 7-branes are encoded in the degeneration of the elliptic fiber over some divisors $\cD_a$ in the base $\cB$ of the fibration $T^2\injto {\cZ}\surjto {\cal B}$. Hence, one expects that the matter zero modes are also encoded in the degeneration of the elliptic fibration. In general the total discriminant $\Delta$ of the elliptic fibration --- which satisfies the D7-brane tadpole condition $[\Delta=0]=12 c_1(\cB)\in H^2(\cB,\IZ)$ --- factorizes as
\beq 
  \Delta=\Delta_{R}\cdot \prod_{a=1}^K \Delta_a^{\delta_a}
\eeq         
where we assume that the vanishing degrees $\delta_a$ are such that, due to the Kodaira, or alternatively the Tate classification (see e.g.~\cite{Bershadsky:1996nh}) of possible fiber degenerations on the {\it smooth} divisor $\cD_a = \{ \Delta_a=0 \} \subset \cB$, we have the gauge group $G_a$. Usually, there remains an $I_1$  rest $\Delta_R$ in the discriminant, which often is not smooth. The simplest example of this is given by a smooth Weierstra\ss\ fibration, where no further gauge enhancements occur and the discriminant has only one component $\Delta=\Delta_R=4\, f^3+ 27\, g^2$. However, the corresponding divisor $\cD_R=\{ \Delta_R=0\} \subset \cB$ is singular over the so-called cusp curve $C_{\rm cusp}=\{f=g=0\}\subset\cD_R$. 

In order to determine the complete matter zero modes for an M5-instanton in F-theory one needs to understand the matter zero modes  coming both from the smooth components $\cD_a$ and from the generically singular component $\cD_R$. It is the aim of this section to study these two kinds of matter zero modes more explicitly.

\subsection[Matter zero modes from smooth \texorpdfstring{$\Delta_a$}{brane}]{Matter zero modes from smooth \boldmath$\Delta_a$}\label{sec_matter}

Let us start with the matter zero modes arising from the intersection of the vertical M5-instanton $\cM=\pi^{-1}(\cE)\subset\cZ$ with the smooth components $\cD_a$ of the discriminant locus supporting a gauge group $G_a$.

To get an idea how this is intrinsically described in F-theory, we consider the simple example of stacks of D7$-$D7' brane pairs in Type IIB and a single $O(1)$ E3-instanton. From the D7-branes we get $U(n_a)$ gauge fields localized on the complex two-dimensional divisors $D_a$ and bifundamental matter fields $\Phi_{\bf (n_a,\ov n_b)}$ localized on the intersection curve of two stacks of D7-branes, i.e.~$C_{ab}=D_a\cap D_b$. The number of these zero modes is given by the dimensions of the cohomology classes
\beq
  H^i\bigl(C_{ab},L_a\otimes L^\vee_b\otimes K_{C_{ab}}^{1/2} \bigr),\quad i=0,1\; .   
\eeq
In addition, Yukawa interactions $\Phi_{\bf (n_a,\ov n_b)} \Phi_{\bf (n_b,\ov n_c)} \Phi_{\bf (n_c,\ov n_a)}$ are localized on the points where three such matter curves meet. 

The E3-instanton now wraps a two-dimensional ``upstairs'' divisor $E\subset {X}$. From the intersection curves $C_{ea}= E\cap D_a$ one gets charged matter zero modes $\lambda_{\bf n_a}$ in the fundamental representation of $U(n_a)$. Their number can be computed from the cohomology classes
\beq\label{matternumber}
  H^i\bigl(C_{ea},L_a \otimes K_{C_{ea}}^{1/2} \bigr),\quad i=0,1\; .   
\eeq
Moreover, Yukawa-type interactions of the form
\beq
  \Phi_{\bf (n_a,\ov n_b)}\, \lambda_{\bf \ov n_a}\, \lambda_{\bf n_b}   
\eeq 
are generated on the triple intersection points $E\cap D_a\cap D_b$.

\begin{figure}[ht]
  \vspace{1.0cm}
  \centering
  \includegraphics[width=0.4\textwidth]{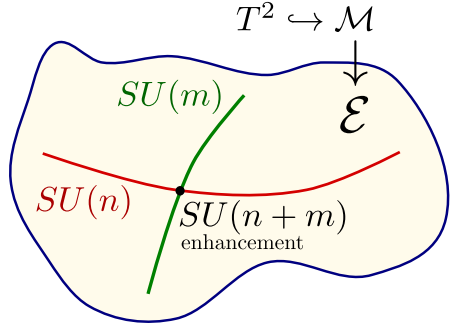}
  \caption{\small $SU(n)$ matter zero mode curves and Yukawa-type interactions in surface $\cE\subset \cM$}
  \label{fig_ethree_enh}
\end{figure}

In the F-theory uplift, a surface $\cD_a$ appears as a component of the discriminant $\Delta=\prod_a \Delta^{n_a}\cdot \Delta_{R}$, i.e.~the elliptic fibration $T^2\injto \cZ\surjto {\cal B}$ degenerates over the surfaces $\cD_a=\{\Delta_a=0\}\subset {\cal B}$ of the type $SU(n_a)$, i.e.~the massive diagonal $U(1)\subset U(n_a)$ has been integrated out. This means that the line bundle $L_a$ is not visible in F-theory. The appearance of matter fields on ${\cal C}_{ab}=\cD_a\cap \cD_b$ can be seen in F-theory as an enhancement of the ADE singularity type over ${\cal C}_{ab}$ to $SU(n_a+n_b)$. Similarly, Yukawa couplings are localized on the points where a further enhancement to $SU(n_a+n_b+n_c)$ happens. The E3-instanton on $E$ lifts up to a vertical M5-brane instanton $\cM=\pi^{-1}(\cE)$, i.e.~$\cM$ is a complex threefold, where $\cE = E/\sigma$ is a complex surface. Clearly, this threefold is by itself elliptically fibered $T^2\injto \cM \surjto \cE$. The matter instanton zero modes $\lambda_{\bf n_a}$ are then localized on the curves $C_{ea}\subset \cE$ where the elliptic fibration develops an $SU(n_a)$ singularity and we expect that their number is still computed by \eqref{matternumber} with $L_a={\cal O}$. This singularity is further enhanced to $SU(n_a+n_b)$ on the points where $\cD_a$ and $\cD_b$ intersect inside $\cE$. The singular geometry just described is schematically depicted in Figure \ref{fig_ethree_enh}.

The way we have just formulated the appearance of M5-instanton matter zero modes and its Yukawa-type interactions intrinsically as degenerations of the elliptically-fibered worldvolume of the M5-brane can be generalized directly to other degenerations of the elliptic fibration. We have listed two straightforward possible local structures in Table \ref{tabledeformcx}.

\begin{table}[ht]
  \centering
  \begin{tabular}{c|c|c|c}
    $G_a$ &  rep.~of $\lambda$    & $G_a\subset \tilde G_b$ & Yukawa   \\ \hline \hline
    $SU(n_a)$      & ${\bf n_a}$  & $SU(n_a)\times SU(n_b)\subset SU(n_a+n_b)$       & $\Phi_{{\bf (n_a,\ov n_b)}}\, \lambda_{\bf \ov n_a}\, \lambda_{\bf n_b}$\uspc \\
    $SO(2n_a)$     & ${\bf 2n_a}$ & $SO(2n_a)\times SU(n_b)\subset SO(2n_a+2\,
    n_b)$ & $\Phi_{\bf (2n_a,\ov n_b)}\, \lambda_{\bf 2n_a}\, 
\lambda_{\bf n_b}$\lspc
  \end{tabular}
  \caption{\small Matter zero modes and Yukawa interactions for  M5-brane instantons in F-theory.}
  \label{tabledeformcx}
\end{table} 

It is now tempting to speculate that the structures just described also apply to  M5-brane instantons intersecting exceptional components of the discriminant locus. To collect  further support for this, let us consider the duality of F-theory on K3-fibered Calabi-Yau fourfolds and the heterotic string on elliptically fibered threefolds.

In this case, the base manifold $\cB$ enjoys a $\IP^1$ fibration over a complex surface ${\cal S}$, i.e.~$\IP^1 \injto \cB\surjto {\cal S}$. Now, let us assume that over ${\cal S}$ we have a degeneration  of the elliptic fiber of the type GUT type $G\in\{SU(5),SO(10),E_6\}$.  As shown in \cite{Hayashi:2009ge,Donagi:2009ra}, we can describe the breaking of $E_8$ to these gauge groups via the spectral cover construction.  Denoting $n={\rm rk}(H)$ with $H=E_8/G$, there exists a spectral surface $C$, which we assume to be  a non-degenerate  $n$-fold cover of the surface ${\cal S}$ carrying a line bundle ${\cal N}$. With the spectral cover $\smash{C \stackrel{\pi_C}{\surjto} {\cal S}}$, the push-forward $V=\pi_{C\, *}\,{\cal N}$ defines an $SU(n)$ bundle over ${\cal S}$. Now let us consider an M5-brane, which is also  vertical with respect to the $\IP^1$ fibration of $\cB$, i.e.~it wraps a complex curve $\Sigma\subset {\cal S}$. 

The dual heterotic model is compactified on the Calabi-Yau threefold $X$ defined via the elliptic fibration $T^2 \injto X \surjto {\cal S}$, which arises via double covering the $\IP^1 = T^2/\IZ_2$ of the F-theory base fibration $\IP^1\injto\cB\surjto{\cal S}$. The breaking of one of the $E_8$ factors to $G$ is achieved by an $SU(n)$ vector bundle $V^X$ on $X$, which according to \cite{Friedman:1997yq} is defined by the Fourier-Mukai transform of precisely the spectral cover data $(C,{\cal N})$. The restriction of the vector bundle $V^X$ to the surface ${\cal S}$ is simply given by 
\beq
  V^X\vert_{\cal S}=\pi_{C\, *}\,{\cal N}=V\; .
\eeq
The vertical M5-brane instanton is mapped by the duality to a world-sheet instanton wrapping the curve $\Sigma$. As being discussed for instance in \cite{Distler:1987ee,Buchbinder:2002ic,Curio:2009wn}, the  left-moving fermionic zero modes of this world-sheet instanton are counted by 
\beq
\label{hetmatternumber}
  H^i(\Sigma, V^X\vert_{\Sigma}\otimes K^{1/2}_\Sigma) = H^i(\Sigma, V\vert_{\Sigma}\otimes K^{1/2}_\Sigma), \quad i=1,2 \; .
\eeq
and transform in the singlet representation of the non-abelian gauge group $G$. This is also what we see in F-theory, as these modes arise from the intersection of the M5-brane with the spectral cover. 

The question is whether there can in principle  be zero modes transforming in non-trivial representation of $G$. Let us consider as an example the case of $G=SO(10)$. Since here one gets matter fields in both the spinor ${\bf 16}$ and the vector ${\bf 10}$ representation of $SO(10)$, one might be tempted to speculate that one should get respective matter zero
modes in these two representations. From our discussion before, the natural guess would be that their numbers are given by
\beq\label{twokindssoten}
  {\bf 16}:\  H^i(\Sigma, K^{1/2}_\Sigma),\qquad
  {\bf 10}:\  H^i(\Sigma,  K^{1/2}_\Sigma) \; .
\eeq
In particular, this means that for $\Sigma=\IP^1$ these zero modes are completely absent and that for higher genus curves they always come in pairs, whose number depends on the spin structure of the curve $\Sigma$. By duality we expect the same formula to hold  in the heterotic theory.

Let us discuss the microscopic origin of these two kinds of matter zero modes in eq.~\eqref{twokindssoten} from the F-theory respectively the Type IIB point of view. Here they arise from the intersection of the instanton with the discriminant locus supporting an $SO(10)$ gauge group. Since the instantonic brane is like a probe brane, there is no extra enhancements of the singularity type of the elliptic fibration over the intersection curve $\Sigma$. What we can use though is the Type IIB description in terms of stacks of $(p,q)$ 7-branes with   string junctions providing the massless modes. Using the conventions of \cite{Gaberdiel:1997mg}, the $SO(10)$ stack is given by the 7-branes $A^5 B  C\,$, where $A$ denotes a D7-brane of type $(p,q)=(1,0)$ and $B C$ is the non-perturbative description of an O7-plane, i.e. $B$ is a $(p,q)=(1,-1)$ 7-brane and $C$ a $(p,q)=(1,1)$ brane.  

\begin{figure}[ht]
  \centering
  \includegraphics[height=3cm]{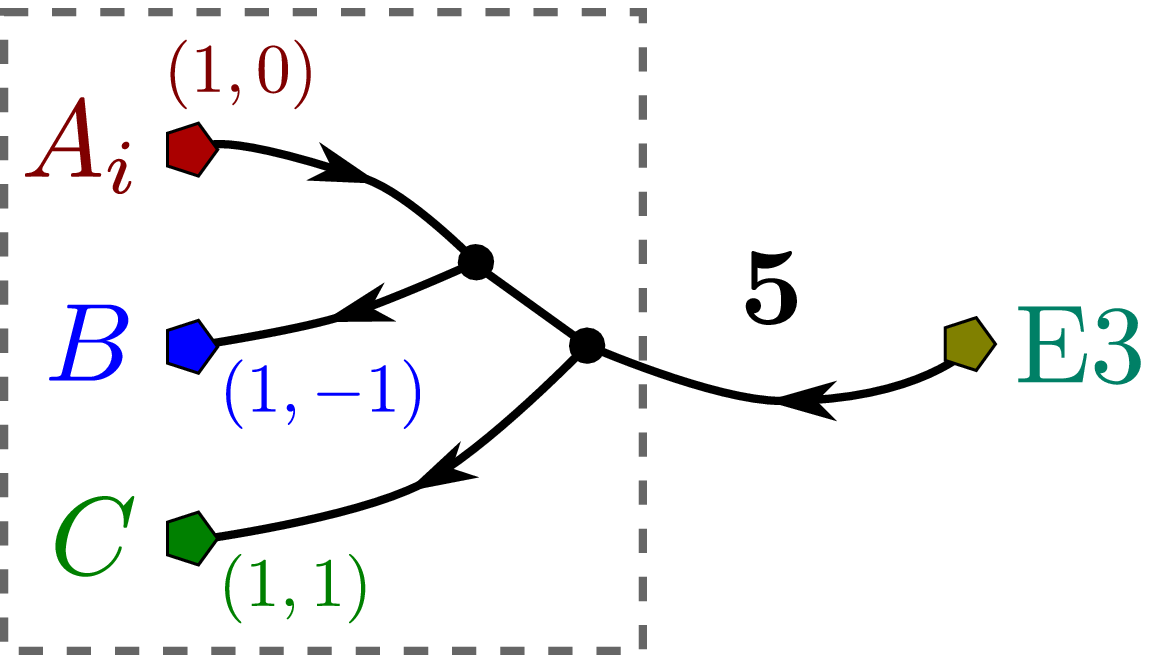}
  \hspace{1.5cm}
  \includegraphics[height=3cm]{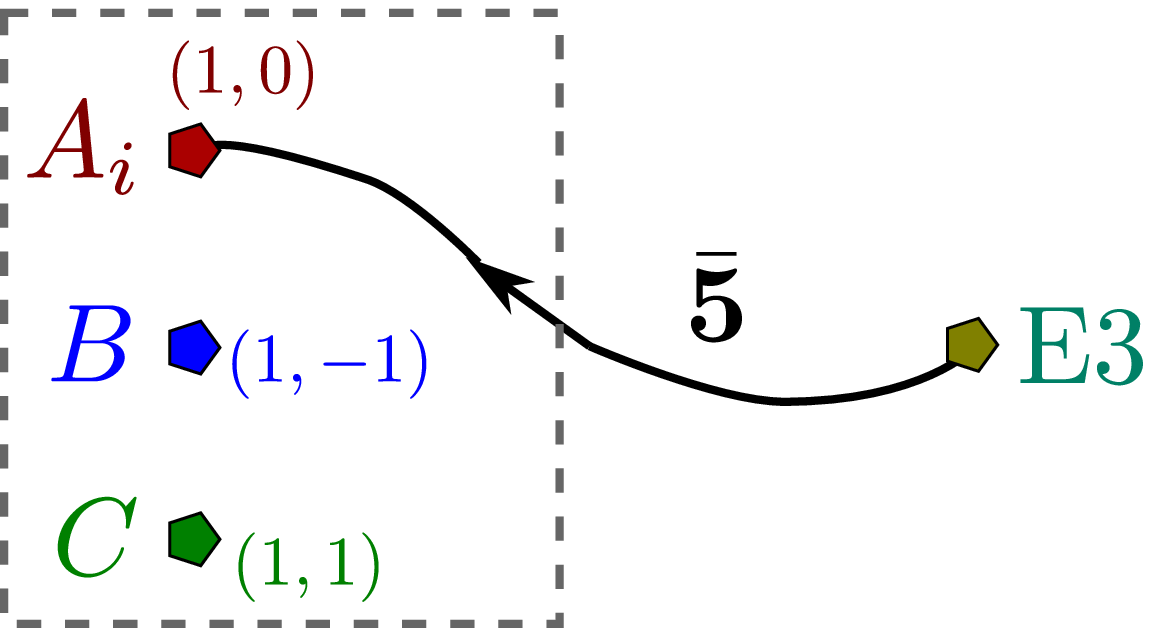}
  \caption{\small String junctions between the 7-brane stack $A^5 BC$ 
       and a single E3-brane
       giving rise to matter zero modes in the ${\bf 10}={\bf 5}+{\bf\ov 5}$
       representation. The ${\bf 5}$ and ${\bf\ov 5}$ is with respect to
    the $SU(5)\subset SO(10)$ on the  five $A$ branes.}
  \label{fig_junctionsten}
\end{figure}

Adding to the $A^5 BC$ configuration a 7-brane of type $A$  gives the enhancement to $SO(12)$, while a brane of type $B$ leads to $E_6$. The massless modes come from various kinds of $(p,q)$ string junctions with their external legs ending on the respective $(p,q)$ 7-branes. The point now is that the E3-brane is $SL(2,\mathbb Z)$ invariant, so that all kinds of $(p,q)$-strings can end on it. Therefore, the external legs from the string junctions extending $SO(10)$ to $SO(12)$ or  $E_6$ can also end on the E3-brane, providing the massless zero modes in the ${\bf 10}$ or ${\bf 16}$ representation of $SO(10)$. For the zero modes in the ${\bf 10}$ representation the resulting picture is shown in figure \ref{fig_junctionsten}

What one might be concerned about now is that for a non-smooth M5-brane divisor $\cM_{\rm sing}$ the number of deformation modes $h^{i,0}(\cM_{\rm sing})$ is first of all not well defined and presumably changed relative to the number $h^{i,0}(\cM_{\rm smth})$ for the smooth Weierstra\ss\ fibration. Indeed to compute $h^{i,0}(\cM_{\rm sing})$ one first has to resolve the ADE singularities of the elliptic  fibration and then compute $\smash{h^{i,0}(\hat \cM)}$ for the resolved $\hat \cM\subset \tilde\cZ$. However, one might expect that by this process only new elements to $\smash{h^{2,1}(\hat \cM)}$ and  $\smash{h^{1,1}(\hat \cM)}$ appear so that we have $\smash{h^{i,0}(\hat \cM)=h^{i,0}(\cM_{\rm smth})}$.

Following our general philosophy of concreteness, let us support this latter claim by a simple though still non-trivial example. Namely, we consider our preferred example of $\IP^4_{11114}[8]$ and its F-theory uplift and do not cancel the 7-brane tadpole by just a single brane but instead split the discriminant into two factors of the form
\beq
  \Delta=\Delta_R\cdot \Delta_2 \xrightarrow{{}\;\;\rm Sen\;\;{}}  h^2 \left(\eta^2_{15} - h\, \chi_{22}\right) u_4^2\, .
\eeq     
This means that on the divisor $u_4=0$ we do have an $SP(2) = SU(2)$ singularity. Since the discriminant has split into two components, we now get two kinds of matter zero modes arising from intersections of the instanton $E_n$ with the $SU(2)$ and $SO(1)$ component of the discriminant, respectively. We will come back to these matter zero modes later, but here we are interested in resolving the $SU(2)$ singularity of the Calabi-Yau fourfold and compute the cohomology classes $\smash{h^{i,0}(\hat \cM)}$ of the resolved M5-brane divisor $\smash{\hat \cM}$.

In \cite{Blumenhagen:2009yv} it was described how to resolve such an $SU(2)$ singularity using methods from toric geometry (see also \cite{Aluffi:2009tm} for similar resolutions). Applying this formalism, for the resolved space we obtain the toric data displayed in Table \ref{tbl:smoothSU2}. The $SU(2)$ singularity is resolved by introducing the new homogeneous coordinate $v$ and a third equivalence relation. From the weights in Table \ref{tbl:smoothSU2} it is clear that the resolved fourfold is not any longer a Weierstra\ss\ fibration. 
\begin{table}[ht]
  \centering
  \begin{tabular}{r@{\,$=$\,(\,}r@{,\;\;}r@{,\;\;}r@{,\;\;}r@{,\;\;}r@{\,)\;\;}|c|ccc|c} 
    \multicolumn{6}{c|}{vertices of the} & coords & \multicolumn{3}{c|}{GLSM charges} & {divisor class}${}^\big.$ \\
    \multicolumn{6}{c|}{polyhedron / fan}      &  & $Q^1$ & $Q^2$ & $Q^3$ &  \\
    \hline\hline
    $\nu_1$ &   1  &   0  &   0  &   0  &   0  & $y$   & 3 & 0 & 1 & $3\sigma+12H-S$\uspc \\
    $\nu_2$ &   0  &   1  &   0  &   0  &   0  & $x$   & 2 & 0 & 1 & $2\sigma+8H-S$  \\
    $\nu_3$ & $-3$ & $-2$ &   0  &   0  &   0  & $z$   & 1 &$-4$&0 & $\sigma$  \\
    $\nu_4$ &$-12$ & $-8$ & $-1$ & $-1$ & $-1$ & $u_1$ & 0 & 1 & 0 & $H$\uspc \\
    $\nu_5$ &   0  &   0  &   1  &   0  &   0  & $u_2$ & 0 & 1 & 0 & $H$  \\
    $\nu_6$ &   0  &   0  &   0  &   1  &   0  & $u_3$ & 0 & 1 & 0 & $H$  \\
    $\nu_6$ &   0  &   0  &   0  &   0  &   1  & $u_4$ & 0 & 1 & 1 & $H-S$  \\
    $\nu_7$ &   1  &   1  &   0  &   0  &   1  & $v$   & 0 & 0 & $-1$ & $S$\lspc  \\ \hline
    \multicolumn{6}{c|}{conditions:}  &                & 6 & 0 & 2 & \uspc \\
  \end{tabular}
  \caption{\small Toric data for elliptically-fibered fourfold over $\IP^3$ with a smooth resolution of the $SU(2)$-enhancement singularity.}
  \label{tbl:smoothSU2}
\end{table}
The Stanley-Reisner ideal of the toric ambient space $\cA$ is 
\beq
  {\rm SR}(\cA)=\langle y x z, \; z v, \; u_1 u_2 u_3 u_4,\; y x u_4, \; u_1 u_2 u_3 v    \rangle
\eeq
which allows one to compute line bundle cohomology classes over the ambient class using \cite{Distler:1996tj, Blumenhagen:1997vt}. The resolved M5-brane $\hat \cM$ is given by a further hypersurface constraint of degree $(0,n,0)$ in the above Calabi-Yau fourfold. It is now a tedious but straightforward exercise to compute 
\beq
  h^{i,0}(\hat \cM)=\Big( 1,0, {\textstyle \binom{n-1}{3}},{\textstyle \binom{n+3}{3}}-1\Big),
\eeq
which agrees with the numbers obtained for the smooth Weierstra\ss\ fibration in \eqref{ftheoryresa}. Therefore, the uncharged zero modes can reliably be computed for the smooth Weierstra\ss\ fibration, and the charged matter zero modes from the intersection of the M5-instanton with various components of the discriminant. Thus, the explicit resolution $\hat\cM$ is not really necessary.

\subsection[Matter zero modes from \texorpdfstring{$\Delta_R$}{remainder}]{Matter zero modes from \boldmath$\Delta_R$}

So far we have described the number of instanton matter zero modes coming from degenerations of the elliptic fiber over smooth curves in the base $\cE$ of $\cM$. However, in most cases the discriminant locus also contains a non-smooth piece $\cD_R$. For the case of a smooth Weierstra\ss\ fibration this comes from the total discriminant $\Delta_R=4f^3+27 g^2$, which has a cusp singularity at $f=g=0$. For computing the total number of matter zero modes, one also has to understand how many extra zero modes come from the singular curve $\cE\cap \cD_R$. In the Type IIB orientifold, these modes arise from the intersection of the E3-instanton on $E$ with the single singular D7-brane $\{\eta^2- \xi^2 \chi=0\}$. In this section, we will compute the expected number of matter zero modes both in the Type IIB orientifold and in the F-theory uplift. For doing this, we first tend to the problem that the  D7-brane in the Type IIB picture has a singular structure (Whitney umbrella), as is explained at length in \cite{Collinucci:2008pf}.

\subsubsection*{Desingularization of the matter curve in the Type IIB picture}
In the Type IIB picture, the matter zero modes residing at the intersection between an $O(1)$ E3-instanton and the singular D7-brane in the ``upstairs'' geometry, are defined by the equations of the form
\beq\label{singcurveiib}
  C: \quad \{\eta^2 = \xi^2\,\chi\} \cap \{ Q_{E}=0\}\,,
\eeq
where $\xi$ is the coordinate that gets reflected by the orientifold involution. This curve has double-point singularities\footnote{Note, that due to the dimensionality of the curve, the so-called pinch-points of the D7-brane at $\eta=\xi=\chi=0$ will never generically lie on the E3-instanton.} at the loci
\beq
  \eta=\xi=Q_{E}=0\,.
\eeq
Our objective is to count the ``matter zero modes'', which are given by $H^0(C, K_C^{1/2})$. 

Let us describe our Calabi-Yau threefold $X$ a bit more precisely. In order to eventually relate this Type IIB setting to F-theory, we need to regard $X$ as the double cover of a threefold $\cB=X/\sigma$ that will be the base of our F-theory fourfold. This is accomplished by taking $\cB$, and adding one more coordinate $\xi$ to it, such that $\xi$ is a section of $K_{\cB}^{-1}$. One now has a fourfold $Y_4$ with coordinates
\beq
  Y_4: \quad (\vec{x}=\underbrace{x_1, \ldots, x_k}_{\cB} , \underbrace{\xi}_{K_{\cB}^{-1}} )\,.
\eeq
To finish building $X\subset Y_4$, we simply impose the hypersurface equation
\beq
  \xi^2=h(\vec{x})\,,
\eeq
such that $X$ is the double covering of $\cB$ branched along the locus $\{h=0\}$ which can be seen as the O7-plane.

In the $X=\IP_{11114}[8]$ model, this construction is how we obtain $X$ from $\IP^3$ by adding a coordinate $\xi$ of degree $4$, and imposing an equation of degree $8$. In the equations \eqref{singcurveiib} for our curve $C$, $(\eta,\, \xi, \, \chi)$ are sections of $(K_{\cB}^{-4},\,  K_{\cB}^{-1}, \, K_{\cB}^{-6})$, respectively, and $Q_{E}$ will be denoted as a section of $N_\cE$, when treated as sections over $\cB$. The point is that we can pull-back all of these bundles onto $X$.

This is a non-transverse system at the locus $\eta=\xi=0$. In order to desingularize the curve $C$, we blow-up the weighted projective ambient space of the Calabi-Yau threefold. After a simplification, this entails adding one more coordinate $t$ to our space, and imposing one more equation. In the end, we are left with the following description of the proper transform $\overline C$ of the curve as a subspace in the fivefold
\beq\label{ambientfivefold}
  Y_5: \quad (\vec{x}=\underbrace{x_1, \ldots, x_k}_{\cB} ;  \underbrace{\xi}_{K_{\cB}^{-1}};  \underbrace{t}_{K_{\cB}^{-3}} )
\eeq
with the constraints
\beq\label{relcurve}
  \overline{C}: \quad \{t\,\xi=\eta\} \cap  \{ \xi^2=h(\vec{x}) \} \cap \{ t^2=\chi \} \cap \{ Q_{E}(\vec{x})=0 \}\,,
\eeq
where the first relation comes from the blow-up constraint, the second is the pullback of the Calabi-Yau hypersurface constraint, the third gives the proper transform of the D7-brane, and the fourth is the pullback of the E3-brane.

\subsubsection*{Counting matter modes in the Type IIB picture}
The canonical bundle of the blown-up curve $\overline{C}$ is easily computed to be
\beq\label{canonicalproperiib}
  K_{\overline{C}} = K_{\cB}^{-7} \otimes N_{E}\,.
\eeq
In order to count the matter modes, we would now like to count sections of the square root of this bundle, \mbox{$K_{\overline{C}}^{1/2} = K_{\cB}^{-7/2} \otimes N_E^{1/2}$}. 

In order to keep track of potentially important information, we will separately count involution-even and -odd sections of this bundle. In order to do so, we need to define a consistent lift of the orientifold involution $\sigma: \xi \mapsto -\xi$ to the auxiliary blow-up space. The choice consistent with the blow-up constraint $t\,\xi = \eta$ is the following involution:
\beq
  \tilde\sigma: \xi \mapsto -\xi\,, \quad t \mapsto -t\,.
\eeq
Clearly, $\tilde\sigma$ projects to the original $\sigma$.

Let us now begin to count elements of $\smash{H^0_+(\ov C,K_{\ov C}^{1/2})}$. Unfortunately, index theorems will not be helpful in this situation, as the indices will be vanishing. The sections must be counted directly. This basically amounts to counting sections of (the pullback) of $\smash{K_{\cB}^{-7/2} \otimes N_E^{1/2}}$ over $Y_5$, made out of the coordinates in \eqref{ambientfivefold}, modulo the ideal\footnote{In this case, we note that the relation $\eta^2-h\,\chi=0$ holds on $\ov C$.} $\mathcal{I}_{\ov C}$ generated by the relations in \eqref{relcurve}.
\beq
  H^0_+(\ov C,K_{\ov C}^{1/2}) = \{ {\rm Sections \ of \ } K_{\cB}^{-7/2} \otimes N_E^{1/2} {\rm \ over \ } Y_5/ \mathcal{I}_{\ov C} \}\,.
\eeq

Involution-invariance means that they can only depend on the combination $\xi t$ or on even powers of $\xi$ and $t$. Fortunately, however, the first three relations in \eqref{relcurve} eliminate all such polynomials in favor of polynomials that depend purely on the $x_i$. Hence, we only need to count sections over $\cB$, modulo the ideal generated by $\langle\eta^2-h\,\chi, \; Q_{E}\rangle$
\beq\label{relposzero}
  H^0_+(\ov C,K_{\ov C}^{1/2}) = \Big\{ {\rm Sections \ of \ } K_{\cB}^{-7/2} \otimes N_E^{1/2}\big\vert_{\cB}\Big/ \langle\eta^2-h\chi, \; Q_{E}\rangle \Big\}\,.
\eeq

Similarly, one can show that $\smash{H^0_-(\ov C,,K_{\ov C}^{1/2})}$ is given by sections over $\cB$ of an appropriate bundle. Since we want involution odd sections, they must be linear in $\xi$ or $t$ (higher powers are eliminated by the relations in \eqref{relcurve}). So we are counting sections of the form $\xi P_{K_{\cB}^{-5/2}}$ and $t P_{K_{\cB}^{-1/2}}$, modulo the ideal $\langle\xi\eta - th, \; t\eta - \xi \chi, \; Q_{E}\rangle$:
\beq\label{relnegzero}
  \bal
    H^0_-(\ov C,K_{\ov C}^{1/2}) &= \Big\{ {\rm Sections \ of \ } K_{\cB}^{-5/2} \otimes N_E^{1/2}\big\vert_{\cB} \Big/ \langle\xi\eta - th, \; t\eta - \xi \chi, \; Q_{E}\rangle \Big\} \\
    & {} \oplus \Big\{ {\rm Sections \ of \ } K_{\cB}^{-1/2} \otimes N_E^{1/2}\big\vert_{\cB} \Big/ \langle\xi\eta - th, \; t\eta - \xi \chi, \; Q_{E}\rangle \Big\}\,.
  \eal
\eeq

So far our discussion has been quite general, but now let us  apply this to our example with $X=\IP_{11114}[8]$ and $\cB=\IP^3$. We remark that the following computation goes through also in the more general case that $\cB$ is a Fano threefold, i.e.~the anti-canonical bundle is ample. For our concrete example  we have $K_{\cB} = -4\,H$ and  we take $Q_{E}=Q_n$ to be a homogeneous equation of degree $n$. In order to be able to treat $\smash{K_{\ov C}^{1/2}}$ as a bundle pulled back from $\cB$, we restrict to even $n$ which gives rise to a divisor $nH$.

To compute $\smash{h^0_+(\ov C, K_{\ov C}^{1/2})}$ we count sections of degree $14+n/2$ over $\cB$, modulo $\langle\eta^2-h\chi,\; Q_n\rangle$. For simplicity, define the following notation:
\beq
  \mathcal{N}(n) = H^0(\cB, {\cal O}(n))=\binom{n+3}{3}\,\theta(n)\qquad {\rm with}\qquad
    \theta(x) = \begin{cases}
    0 & \textnormal{if $x<0$}, \\
    1 & \textnormal{if $x \geq 0$} \; .
  \end{cases} 
\eeq
This function counts the polynomials of degree $n$ in four variables. Then we can compute our dimension as follows:
\beq
\label{sectionzwoba}
  h^0_+(\ov C, K_{\ov C}^{1/2}) = \mathcal{N}(14+\tfrac{n}{2})-\mathcal{N}(14-\tfrac{n}{2})-\mathcal{N}(\tfrac{n}{2}-18)
\eeq
where the last term can be written as $\mathcal{N}(14+\tfrac{n}{2}-32)$ revealing that it  is due to the degree-32 relation $\eta^2-h\,\chi$. After simplifications, it can be shown that
\beq
  h^0_+(\ov C,K_{\ov C}^{1/2}) = \begin{cases}
    \frac{1}{24}\,n\,\left(n^2+3068\right)\, & \textnormal{if $0\leq n < 36$}, \\
    4\,(n^2+340)\, & \textnormal{if $n\geq 36$}\; .
  \end{cases}
\eeq
Now let us count the odd sections. These are simply given by the polynomials of degree $14+n/2$ that are linear in $\xi$ or $t$ (but not both) modulo the $Q_{n}$ relations in \eqref{relnegzero}. Performing a similar computation as before, we can write
\beq
  \bal
    h^0_-(\ov C, K_{\ov C}^{1/2})  &{}= \mathcal{N}(10+\tfrac{n}{2}) - \mathcal{N}(10-\tfrac{n}{2}) -\mathcal{N}(\tfrac{n}{2}-6) \\
    &{}+\mathcal{N}(2+\tfrac{n}{2}) 
    -\mathcal{N}(2-\tfrac{n}{2})-\mathcal{N}(\tfrac{n}{2}-14)\; .
  \eal
\eeq
This can be written as follows:
\beq\label{antisymsecto}
  h^0_-(\ov C,K_{\ov C}^{1/2}) = \begin{cases}
    \frac{1}{12}\,n\,\left(n^2+956\right)\, & \textnormal{if $0\leq n \leq 4$}, \\
    \frac{1}{16}\,\left(n^3+8\,n^2+1212\,n+160\right)\, & \textnormal{if $4< n < 12$}, \\
    \tfrac{1}{24}\,n^3+n^2+\frac{431}{6}\,n+20\, & \textnormal{if $12\leq n \leq 20$}, \\
    \frac{1}{48}\,\left(n^3+120\,n^2+1724\,n+14688\right)\, & \textnormal{if $20 < n < 28$}, \\
    4\,\left(n^2+148\right)\, & \textnormal{if $28 \leq n$}.
  \end{cases}
\eeq
This completes the computation of the matter zero modes arising from the singular component of the discriminant in the Type IIB perspective. One might be concerned that maybe only the invariant or anti-invariant matter zero modes survive the orientifold projection. From the open string perspective, since an $E3\to D7$ open string is mapped to a $D7\to E3$ string, this is actually not expected and one should get $\smash{h^0_+(\ov C,K_{\ov C}^{1/2})+h^0_-(\ov C,K_{\ov C}^{1/2})}$ for the total number of zero modes.

To further support this,  let us also compute the number of charged zero modes for the case that the single D7-brane splits into a brane-anti-brane pair $\eta_{16}\pm\xi \psi_{12}$ so that it carries a $U(1)$ gauge group. As we have seen in section \ref{sec_31}, in this case the two branes carry line bundles $L_\pm={\cal O}(\pm 6)$. Since now the world-volume of the D7-brane is smooth, we can directly compute $\smash{H^i(\Sigma, L\otimes K_\Sigma^{1/2})}$ for $i=0,1$. The curve $\Sigma$ is defined via the two  constraints 
\beq\label{relvvcurve}
  {\Sigma}: \quad \{\eta_{16}+\xi \psi_{12}=0 \} \cap \{ Q_{E}(\vec{x})=0 \}\,,
\eeq
in $\IP_{11114}[8]$. Using Serre duality one eventually has to count the sections
\beq 
  H^0\left(\Sigma, {\cal O}({\textstyle \frac{n}{2}}+14)\right)+
  H^0(\Sigma, {\cal O}\left({\textstyle \frac{n}{2}}+2)\right) \; .
\eeq
Using the same methods as before, one readily obtains
\beq
  \bal
    h ^0\left(\Sigma, {\cal O}({\textstyle \frac{n}{2}}+14)\right)&=
    \mathcal{N}(14+\tfrac{n}{2}) - \mathcal{N}(14-\tfrac{n}{2})
    - \mathcal{N}(\tfrac{n}{2}-2) \\
    &+\mathcal{N}(10+\tfrac{n}{2}) - \mathcal{N}(10-\tfrac{n}{2})
    - \mathcal{N}(\tfrac{n}{2}-6)
  \eal
\eeq
and
\beq
  \bal
    h^0\left(\Sigma, {\cal O}({\textstyle \frac{n}{2}}+2)\right)&=
    \mathcal{N}(2+\tfrac{n}{2}) - \mathcal{N}(2-\tfrac{n}{2})
    - \mathcal{N}(\tfrac{n}{2}-14)  \\
    &+\mathcal{N}(\tfrac{n}{2}-2) 
    - \mathcal{N}(\tfrac{n}{2}-18)\; .
  \eal
\eeq
Adding these two contributions, the term $\mathcal{N}(\tfrac{n}{2}-2)$ drops out and one obtains the same total number of charged zero modes as for the single $O(1)$ D7-brane, i.e.
\beq
  \underbrace{h^0(\Sigma,L\otimes K_\Sigma^{1/2})+h^0(\Sigma,L^\vee\otimes
    K_\Sigma^{1/2})}_{\rm brane/image-brane}=
  \underbrace{h_+^0(\ov C,  K_{\ov C}^{1/2})+h_-^0(\ov C,  K_{\ov
      C}^{1/2})}_{\rm recombined\ brane}\; .
\eeq
We would like to interpret this consistent result as evidence that performing the blow-up of the singular curve $C\to \ov C$ was the mathematically correct procedure to reliably compute the number of matter zero modes. Moreover, the brane/image-brane recombination apparently does not induce any pairing-up of the charged zero modes and indeed in the recombined phase the total number of zero modes are counted by $h_+^0(\ov C,  K_{\ov C}^{1/2})+h_-^0(\ov C,  K_{\ov C}^{1/2})$.

\subsubsection*{Desingularization of the matter curve in F-theory}
Let us now perform a similar computation in F-theory. Moving away from the Sen limit the discriminant does not factorize any longer and is just given by the single factor $\Delta=f^3-g^2=0$. Therefore, the matter zero modes reside on the intersection curve of the  generic discriminant with the E3-instanton given by:
\beq
  \cC_F: \quad \{\Delta=f^3-g^2=0 \} \cap \{ Q_{\cal E}=0 \}
\eeq
where $f$, $g$ and $Q_{\cal E}$ are sections of $K_{\cB}^{-4}$, $K_{\cB}^{-6}$ and $N_{\cal E}$, respectively. This curve has cusp singularities at the loci $f=g=Q_{\cal E}=0$. In order to count the zero modes, we will first desingularize this (non-perturbative) by blowing up the cusp points. After some simplification, we are left with a new ambient fourfold $\cX_4$, whose coordinates are those of $\cB$ plus one more coordinate, which is a section of $K_{\cB}^{-2}$:
\beq\label{ambientfourfold}
  \cX_4: \quad (\vec{x}=\underbrace{x_1, \ldots, x_k}_{\cB}, \underbrace{s}_{K_{\cB}^{-2}})\,.
\eeq
The resolved curve $\overline{\cC}_F$ is then defined by the constraints
\beq\label{relcurveF}
  \overline{\cC}_F: \quad \{ s^2=f \} \cap \{  s^3=g \} \cap \{ Q_{\cal E}(\vec{u})=0 \}\,,
\eeq
whereby the first two relations are equivalent to the blow-up constraint plus the proper transform constraint. For our standard example $\IP_{11114}[8]$, this can be written as the complete intersection of three hypersurface constraints in $\cX_4$ as shown in Table \ref{tableresos}.

\begin{table}[ht]
  \centering
  \begin{tabular}{c c c||ccccc}
    $Q$ & $F$ & $G$   & $u_1$ & $u_2$ & $u_3$ & $u_4$ & $s$ \\ \hline
    $n$ & $16$ & $24$ & 1 & 1 & 1 & 1 & 8\uspc
  \end{tabular}
  \caption{GLSM data of the resolved cusp curve.}
  \label{tableresos}
\end{table}

\subsubsection*{Counting matter modes}
From the data in the previous section we can compute that
\beq
  K_{\overline{\cC}_F} = K_{\cB}^{-7}\otimes N_{\cE}\,.
\eeq
Remarkably, this is the same bundle as $K_{\overline{C}}$ in the weak coupling limit \eqref{canonicalproperiib}.

\begin{figure}[ht]
  \begin{center}
    \includegraphics[width=0.6\textwidth]{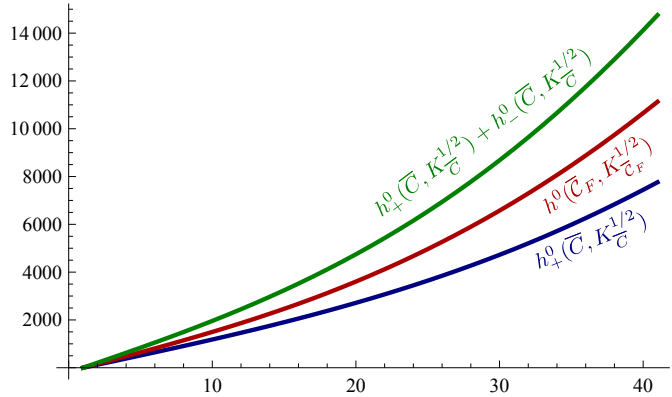}
  \end{center}
  \caption{\small The three moduli functions show the relation of \eqref{eq_modulirelations}.}
  \label{fig_moduli}
\end{figure}

Let us now proceed to count sections of $\smash{K_{\overline{\cC}_F}^{1/2}}$. The constraint $s^2=f$ can be used to eliminate all sections of degree two and higher in $s$. The number of these sections would be given by
\beq\label{ftheorysectiona}
  \bal
    h^0_{s-{\rm indep}}(\overline{\cC}_F,K^{1/2}_{\overline{\cC}_F})&=\mathcal{N}(14+\tfrac{n}{2}) - \mathcal{N}(14-\tfrac{n}{2}) - \mathcal{N}(\tfrac{n}{2}-34)  \\
    h^0_{s-{\rm linear}}(\overline{\cC}_F,K^{1/2}_{\overline{\cC}_F})&=\mathcal{N}(6+\tfrac{n}{2}) - \mathcal{N}(6-\tfrac{n}{2})\; .
  \eal
\eeq
Next we have to take care of the constraint $s^3=s\, f= g$, which  relates $s$-independent sections to $s$-linear sections. Eliminating $s$-linear sections by $s$-independent sections by $\{s\, f =g\},\{s\, g=f^2\}$ and repairing that  we have double counted sections $s\, f\, g = g^2$, we have to add to \eqref{ftheorysectiona} the three contributions
\beq
  \delta h^0(\overline{\cC}_F,K^{1/2}_{\overline{\cC}_F})= -\mathcal{N}(\tfrac{n}{2}-10) - \mathcal{N}(\tfrac{n}{2}-18) + \mathcal{N}(\tfrac{n}{2}-34)  
\eeq
so that in total we find
\beq\label{ftheorysectionb}
  \bal
    h^0(\overline{\cC}_F,K^{1/2}_{\overline{\cC}_F})={}&\mathcal{N}(14+\tfrac{n}{2}) - \mathcal{N}(14-\tfrac{n}{2}) -\mathcal{N}(\tfrac{n}{2}-18)\\
    &+\mathcal{N}(6+\tfrac{n}{2}) - \mathcal{N}(6-\tfrac{n}{2}) -\mathcal{N}(\tfrac{n}{2}-10)\; .
  \eal
\eeq
We observe that this result is different from the Type IIB result but that the first line in \eqref{ftheorysectionb} corresponds precisely to the result for $\smash{h_+^0(\ov C,  K_{\ov C}^{1/2})}$ in \eqref{sectionzwoba}. As shown in Figure \ref{fig_moduli}, we have the relation
\beq\label{eq_modulirelations}
  \underbrace{h_+^0(\ov C,  K_{\ov C}^{1/2})}_{\rm IIB\ orient.}
  \ <\  \underbrace{h^0(\overline{\cC}_F,K^{1/2}_{\overline{\cC}_F})}_{\rm F-theory}\  <\ 
  \underbrace{h_+^0(\ov C,  K_{\ov C}^{1/2})+h_-^0(\ov C,  K_{\ov C}^{1/2})}_{\rm IIB\ orient.}\ .
\eeq
Assuming that blowing up the cusp curve is physically the right way to proceed, our results indicate that going away from the perturbative Type IIB orientifold limit, some (but not all) of the matter zero modes are paired up.

\begin{table}[ht]
  \centering
  \begin{tabular}{c c ||cccc}
  $Q$ & $R$ & $u_1$ & $u_2$ & $u_3$ & $u_4$ \\ \hline
  $n$ & $16$    & 1 & 1 & 1 & 1\uspc
  \end{tabular}
  \caption{GLSM data of the auxiliary curve $\cC_A$.}
  \label{tableCA}
\end{table}

Finally we observe that the number of zero modes on the F-theory side can be represented as
\beq
  h^0(\overline{\cC}_F, K_{\overline{\cC}_F}^{1/2}) = h_+^0(\overline{C},K_{\overline{C}}^{1/2}) + h^0(\cC_A, K_{\cC_A}^{1/2})
\eeq
if an additional curve $\cC_A$ is introduced, which is specified by the toric data in Table \ref{tableCA}. This curve could be described by the intersection of the vertical M5-brane divisor with a factor of the form $\,\epsilon P_{16}-2h^2$ in the discriminant, which can be related to the O7-plane in the Sen limit $\epsilon\to 0$. With the so far collected information, it is not clear whether this is just a coincidence or whether this tells us that the actual number of F-theory matter zero modes is given by deducting this contribution from the naive F-theory result. In this later case, the F-theory matter zero modes would be given by $\smash{h^0_+(\overline{C},K_{\overline{C}}^{1/2})}$. Hopefully, future study will clarify this subtle point.

\subsection{Phenomenological consequences}

The results of this section show that also in F-theory there exist charged matter zero modes, which are not included in $\chi({\cal M},{\cal O}_{\cal M})$ and which carefully have to be taken into account for deciding to which four-dimensional effective couplings an M5-brane instanton can make a correction. Generically, an M5-brane will intersect certain components of the discriminant locus leading to matter zero modes in the fundamental representation of the respective gauge group. This included also any singular component --- generically a non-ADE degeneration of Kodaira type $I_1$ --- of the discriminant. Therefore, when building for instance concrete global F-theory realizations of GUT models, only a careful zero mode analysis can tell whether there are non-perturbative effects which can for instance break supersymmetry via a Polonyi-type mechanism or induce KKLT type instanton corrections to the closed string superpotential. 

Indeed, for moduli stabilization respectively for deciding whether an ${\cal N}=1$ string compactification is non-perturbatively unstable, one needs to look for instantons which contain no charged matter zero modes at all. A sufficient condition for the existence of such a potentially destabilizing instanton is that in addition to the M5-brane being rigid, i.e.~$h^{i,0}(\cM)=0$ for $i\in\{1,2,3\}$, all intersections of the M5-instanton with components $\cD_a$ of the discriminant occurs over curves of genus zero such and that the bundle $V_a$ restricts trivially on it. Indeed, only in this case one finds
\beq \label{secondcriterion}
  \cE\cap \cD_a\cong\IP^1 \quad \leadsto \quad   H^*\big(\IP^1,{\cal O}(-1)\big)=(0,0)\; .
\eeq
This is directly analogous to the known heterotic string condition \cite{Distler:1987ee}, that a world-sheet instanton on a rational curve can only contribute to the uncharged superpotential, if the vector-bundle restricts trivially on it.

One should keep in mind that what we have discussed so far is just the zero mode structure of M5-brane instantons. The complete instanton generated amplitude does not only involve the integral over the zero modes and its absorption amplitudes, but --- at least for holomorphic ${\cal N}=1$ couplings like the superpotential or the gauge kinetic function --- a one-loop determinant for the fluctuations around the instanton solution. Recall that in the Type IIB compactification, this one-loop determinant was nothing else than just the exponential of the open string one-loop amplitudes, i.e.~the annuli and M\"obius strip amplitudes, involving at least one boundary on the E3-instantonic brane. Therefore, one expects that also in F-theory the corresponding one-loop determinant receives contributions from open Euclidean membranes ending on the M5-brane.

In the next sections, we will elaborate on the possible zeroes of the one-loop determinant.

\subsection{Counting three-cycles on the M5-brane}

As stated in the previous section, counting zero modes is not sufficient to determine the structure of the effective term generated by an instanton. One must also have information about the one-loop determinant. Since this determinant is typically a section of a line bundle over a (complex structure, and open string) moduli space, it will have vanishing loci. 

The structure of the one-loop determinant was elucidated from the M5-brane point of view in \cite{Witten:1996hc}, where it was shown that the presence of three-cycles in the M5-divisor can lead to cancellations. Therefore, we will now proceed to study the three-cycles of the M5-brane.

Intuitively it is clear that there should be a relation\footnote{In \cite{Diaconescu:1998ua} such a correspondence was found for local colliding $SU(n_c)\times SU(N_f)$ singularities with a (gauge)  instanton on the colour $SU(n_c)$.} between the topology of the F-theory M5-brane divisor $\cM=\pi^{-1}(\cE)$ and the resolved Type IIB intersection curve $\ov C$ of the single D7-brane and the instanton $E$. More specifically, one should be able to determine the topology of $\cM$ by knowing the topology of the double cover $E$ of its base $\cE$, and how the elliptic fiber degenerates over it. We have already seen that involution-even cycles in $E$ become cycles of $\cM$. We have also seen that an involution-odd $n$-cycle will combine with the $A$-cycle of the elliptic fiber (which is itself involution-odd), into an $(n+1)$-cycle of $\cM$. But we have not yet accounted for the full Hodge diamond of $\cM$. There are some cycles in $\cM$ that do not originate from cycles in $E$. Let us reveal this relation for our concrete example.

The aim now is not just to compute $h^{i,0}(\cM)$ but the whole Hodge diamond (see \cite{Braun:2009bh} for related  work on Calabi-Yau threefolds), which means in particular $h^{1,1}(\cM)$ and $h^{2,1}(\cM)$. For doing this we proceed by first computing the Euler characteristic of $\cM$
\beq\label{eulerweier}
  \chi(\cM)=-48\,n\,(n+20)\; ,
\eeq
so that using the already known $h^{i,j}(\cM)$ we can immediately conclude
\beq
  h^{2,1}(\cM)-h^{1,1}(\cM)= \frac{n^3}{6} +21\,n^2 +\frac{2891\, n}{6} -1\; .
\eeq
We expect that $h^{1,1}(\cM)=h_+^{1,1}(E)+1$. For the orientifold invariant and anti-invariant elements in $h^{1,1}(E)$ we find
\beq
  h^{1,1}_+(E)=\frac{2\, n^3}{3}  -2\, n^2 +\frac{7\,n}{3}, \qquad
  h^{1,1}_-(E)=\frac{2\, n^3}{3}  +2\, n^2 +\frac{103\, n}{3},
\eeq
which is consistent with $\chi(E)=2n\,(n^2+22)$. Therefore, we obtain for the M5-brane divisor
\beq
  \bal
    h^{2,1}(\cM)&=\frac{5\, n^3}{6}  +19\, n^2 +\frac{2905\, n}{6}\\
    h^{1,1}(\cM)&=\frac{2\, n^3}{3}  -2\, n^2 +\frac{7\, n}{3} +1 \; .
  \eal
\eeq
Except for the elliptic fiber, all K\"ahler classes of $\cM$ are inherited from the invariant K\"ahler classes on $E$. As before, the anti-invariant elements $H^{2,0}_-(E)$, $H^{1,1}_-(E)$ and  $H^{0,2}_-(E)$ give rise to elements in the third cohomology of $\cM$
\beq
  \begin{array}{l}
          H^{2,0}_-(E) \to  {\begin{cases} H^{3,0}(\cM) \\ H^{2,1}(\cM) \end{cases}}_\big. \\
    H^{0,2}_-(E) \to  {\begin{cases} H^{1,2}(\cM) \\ H^{0,3}(\cM)\end{cases}}^\big.
        \end{array} \quad
        H^{1,1}_-(E) \to  \begin{cases} H^{2,1}(\cM) \\ H^{1,2}(\cM) \end{cases}.
\eeq
We have already seen that actually all elements of $H^{3,0}(\cM)$ come from $H^{2,0}_-(E)$. However, not all elements of $H^{2,1}(\cM)$ can be explained by elements of $H^{2,0}_-(E)$ and $H^{1,1}_-(E)$. Indeed there are precisely
\beq
  \hat h^{2,1}(\cM)=h^{2,1}(\cM)-h_-^{1,1}(E)-h_-^{2,0}(E)=16\,n\, (n+28)
\eeq
such extra elements. Observing that the factor $(n+28)$ looks familiar from eq.~\eqref{canonicalproperiib}, we realize that $\hat h^{2,1}(\cM)$ is directly related to the Euler characteristic of the curve $\ov C$ in the Type IIB orientifold
\beq\label{1cyclesto3cycles}
  \hat h^{2,1}(\cM) + \hat h^{1,2}(\cM) = -\frac{1}{2}\, \chi(\ov C) = {\ov g -1 }
\eeq
where $\ov g$ is the genus of the curve $\ov C$. This result can be understood as follows. Every non-trivial one-cycle $H_1(\ov C)$ of the Riemannian surface $\ov C$ is actually trivial as a one-cycle in the surface $E$, since $H_1(E)=0$. This means that each one-cycle $Z_1 \in H_1(\ov C)$, can be written as the boundary of some two-chain $C_2 \subset E$
\beq
  Z_1 = \partial C_2\,.
\eeq
Let us now look at how this chain comes about in  the geometry of the M5-brane. Choose a basis $A,B$ for the one-cycles $H_1(T^2,\IZ)$ of the elliptic fiber, such that the $A$-cycle collapses in the presence of a (perturbative) D7-brane. We notice that the $A$-cycle can be fibered over the two-chain $C_2$, and that, by definition, it will collapse above the locus $\partial C_2 = Z_1$, since the two-chain ends on a D7-brane. This means that the $A$-cycle fibration over $C_2$ is a three-cycle $Z_3 \in H_3(\cM)$. In other words, one-cycles at the (desingularized) intersection curve $\ov C$ of $\mathrm{E3} \cap \mathrm{D7}$ give rise to three-cycles in the M5-brane.

Due to the orientifold projection, however, not all one-cycles will give rise to a three-cycle. The $A$-cycle in the elliptic fibers suffers the monodromy $A \mapsto -A$ under the orientifold operation. Therefore, in order to make a three-cycle that survives the combined orientifold actions on the base space and the elliptic fiber, only involution-odd two-chains, $\sigma^* C_2 =-C_2$, can be used. Hence, only \emph{involution-odd one-cycles in $\ov C$ give rise to three-cycles in $\cM$}. Since all elements in $H^{0,3}(\cM)$ have been accounted for in our analysis, we conclude that 
\beq
  \hat h^{2,1}(\cM) + \hat h^{1,2}(\cM) = b^1_-(\ov C)\,.
\eeq
The Lefschetz fixed-point theorem tells us that 
\beq
  b^1_-(\ov C) = \frac{\chi(\ov C^{\tilde\sigma})-\chi(\ov C)}{2}\,.
\eeq
where $\ov C^{\tilde\sigma}$ are the fixed points in $\ov C$ under the involution $\tilde \sigma: (\xi, t) \mapsto (-\xi, -t)$. By inspecting \eqref{relcurve}, we see that $\smash{\ov C^{\tilde\sigma} = \emptyset}$, from which we conclude that 
\beq
  b^1_-(\ov C) = -\frac{1}{2}\,\chi(\ov C)\,,
\eeq
exactly as predicted in \eqref{1cyclesto3cycles}.

It is in this sense that the curve on which the matter zero modes are localized is encoded on the geometry of the F-theory divisor $\cM$. Therefore, at least for the case of just a single D7-brane in Type IIB, i.e.~a smooth Weierstra\ss\ fibration of the Calabi-Yau fourfold, we have related the whole Hodge diamond of the M5-brane threefold divisor $\cM$ to the (equivariant) cohomology of the Type IIB $O(1)$ E3-brane divisor $E$ and its intersection with the D7-brane. The lower half of the Hodge diamond of $\cM$ can be written as

\begin{table}[ht]
  \[
    \bal
      & \hspace{5.8cm} 1 \\[0.4cm]
      & \hspace{3.5cm} h^{1,0}_+(E) \hspace{2cm}    h^{0,1}_+(E) \\[0.4cm]
      & h^{2,0}_+(E)+h^{1,0}_-(E) \hspace{1.8cm} h_+^{1,1}(E)+1 \hspace{1.9cm}
        h^{0,2}_+(E)+h^{0,1}_-(E) \\[0.4cm]
        h^{2,0}_-(E)  &\hspace{1.0cm}  h^{2,0}_-(E)+h^{1,1}_-(E)+{\textstyle
          \frac{\ov g -1}{2}}
        \hspace{1.0cm} h^{0,2}_-(E)+h^{1,1}_-(E)+{\textstyle \frac{\ov g -1}{2}}  \hspace{1.0cm} 
        h^{0,2}_-(E) 
    \eal
  \]
  \caption{Hodge diamond of the M5-brane.}
        \label{tab_hodgediamond}
\end{table}

We expect this relation between three-cycles in the M5-brane divisor and one-cycle in the matter zero mode curves to generalize to the case when the discriminant splits into more components over which the elliptic fiber can degenerate. To provide an example of this we consider the discriminant already discussed in the section \ref{sec_matter}, i.e.~the split into an $SU(2)$ locus and a remaining $SO(1)$ locus. In this case, we find for the Euler characteristic of the resolved M5-brane divisor $\hat \cM$ in the fourfold geometry in Table \ref{tbl:smoothSU2} 
\beq
  \chi(\hat \cM)=-48\, n^2 -846\, n
\eeq
which compared to \eqref{eulerweier} shows that the Euler characteristic has increased relative to the Weierstra\ss\ fibration. This means that the number of three-cycles has decreased. We have already shown that $\smash{h^{i,0}(\hat \cM)}$ does not change relative to the smooth Weierstra\ss\ fibration. Due to the resolution of the $SU(2)$ singularity, we expect that the number of K\"ahler forms increases by one, i.e.~$h^{1,1}(\hat \cM)=h^{1,1}( \cM)+1$, which using $\chi(\hat\cM)$ allows to compute $h^{2,1}(\hat\cM)$ as well. From this we can conclude that the contribution to $\smash{b^3(\hat \cM)}$ not yet accounted for is
\beq\label{resolvedthree}
  \hat h^{2,1}(\hat \cM)+\hat h^{1,2}(\hat \cM)=2\, (16\, n^2 +391\, n +1)\; .
\eeq
These are the three-cycles, we expect to arise from the matter zero mode curves $C_1$ and $C_2$. The contribution from the intersection with the $SO(1)$ divisor $\eta^2_{15}+ h\, \chi_{22}$ can be computed as before
\beq
  \hat h^{2,1}(\hat \cM)+\hat h^{1,2}(\hat \cM)\vert_{SO(1)}= b^1_-(\ov C_1)= -\frac{1}{2}\chi(\ov C_1)= 30\, n\, (n+26)\; .
\eeq
The Euler-characteristic of the $SU(2)$ matter curve $C_2$ is $\chi(C_2)=-2n(n+1)$, i.e.~the curve $C_2$ has $\ 2g(C_2)=2n(n+1)+2\ $ one-cycles. If each of these one-cycles induces a three-cycle in $\hat \cM$ we get
\beq
  \hat h^{2,1}(\hat \cM)+\hat h^{1,2}(\hat \cM)\vert_{SU(2)}=2\, g(C_2)= 2\, n\, (n+1)+2\; .
\eeq
Adding the contributions from the two matter curves $C_1$ and $C_2$ we precisely get the number \eqref{resolvedthree} expected for the resolved M5-brane divisor.

\subsection{Three-cycles and the M5 partition function}

In \cite{Witten:1996hc}, the partition function of the M5-brane was studied from many points of view. The important result was that, beside the presence of the zero modes on which we have focused so far, the partition function of the M5-brane, and hence the putative effective superpotential that it can generate, may suffer from cancellations if the M5-brane divisor has a non-trivial third cohomology group $H^3(\cM)$. 

More specifically, the effective theory of the M5-brane is known to contain a ``chiral'' two-form $\beta_2$ propagating on its worldvolume, whose field-strength $T_3$ is self-dual. The M5-brane effective action also couples to the bulk 11d supergravity threeform $C_3$, of field-strength $G_4$, which, in the analysis of \cite{Witten:1996hc}, is subject to the condition $G_4\vert_{\cM} =dT_3$. This implies that the restriction of $G_4$ to the M5-brane must be cohomologically trivial. The moduli space of ``flat three-form connections'' $C_3$ restricted to $\cM$ is a torus, known as the ``intermediate Jacobian'' of $\cM$:
\begin{equation}
J_{\cM} = H^3(\cM, \mathbb{R})/H^3(\cM, \mathbb{Z})\,.
\end{equation}
Witten showed that the partition function of the worldvolume two-form $\beta_2$ is a section of a holomorphic line bundle ${\cal L}$ over the intermediate Jacobian, $J_{\cM}$, of $\cM$:
\begin{equation*}
Z(\beta_2) \in \Gamma({\cal L}) \quad {\rm with } \quad \pi: {\cal L} \rightarrow J_{\cM}\,.
\end{equation*}
By virtue of being the section of a line bundle, as opposed to a function, the partition function will vanish on (real) codimension two loci over the moduli space of the supergravity $C_3$ form. In the models of \cite{Diaconescu:1998ua}, for instance, these vanishing loci were interpreted as points in the $C_3$ moduli space where quark-like zero modes become massless.
\vskip 2mm
Armed with our dictionary for the Hodge diamond of $\cM$ in terms of the equivariant Hodge diamond of E3-divisor $E$ --- see Table \ref{tab_hodgediamond} --- we can now shed some light on this intermediate Jacobian from the perturbative, Type IIB orientifold point of view. The intermediate Jacobian can be written in the following factorized form:
\begin{equation}
 J_{\cM} = \left(H^2_-(E, \mathbb{R})\right)^2 \times H^1_-(\ov C,\mathbb{R})/ \left(\left(H^2_-(E, \mathbb{Z})\right)^2 \times H^1_-(\ov C,\mathbb{Z})\right)\,.
 \end{equation}
 We can now isolate the individual contributions and interpret them. The contribution from $H^1(\ov C)$ is the expected contribution from the matter zero modes emanating from the D7/E3 intersection curve $\ov C$. Here we see that, only when this (desingularized) intersection curve has genus zero, does the contribution to $J_{\cM}$ vanish.
 
The $H^2_-$ factor contains the invariant geometric moduli of the instanton $h^{0,2}_-$, which we already expect to be hazardous to the generation of the superpotential. However, it also has a contribution from the $h^{1,1}_-$ Hodge number, which is a priori not detected by the standard zero mode counting of instantons. This contribution has the following interpretation. The dimensional reduction of the supergravity threeform $C_3$, whose moduli space we are parametrizing, gives rise to the NSNS $B_2$-field in ten dimensions. The orientifold projection requires that $B_2$ be involution odd. In orientifold models with $b^{2}_-(X_3)=0$, such as our favored $\mathbb{P}^4_{11114}[8]$ example, this field is a priori projected out. However, if the cohomology of the E3-instanton has $b^2_-(E) \neq 0$, then the $B_2$-field is allowed to take on configurations that are non-trivial in the cohomology of the E3, even though they are trivial in the Calabi-Yau threefold. It is precisely these $B_2$-field moduli that contribute to $J_{\cM}$.

Hence, even if the E3-instanton satisfies the criterion that $h^{0,i}(\cM)=0$ \emph{and} our criterion \eqref{secondcriterion}, its superpotential can still have zeroes at non-generic loci in the closed string moduli space. Hence, an even stronger, sufficient criterion to have a nowhere vanishing, uncharged superpotential is to impose, in addition, that 
\begin{equation} \label{thirdcriterion}
b^2_-(E) = 0\,.
\end{equation}

\section{Conclusion}

In this paper we have considered the relation between E3-brane instantons in Type IIB orientifold settings and M5-brane instantons in F-theory. Here we have put special emphasis on the appearance of the various kinds to bosonic and fermionic zero modes. In Type IIB string theory these can be described as Euclidean open strings with at least one boundary on the E3-brane, whereas in F-theory they are given by open membranes with at least one two-dimensional boundary on the M5-brane. Instead of quantizing this latter system from first principles, we have gained quite a lot of information by carefully uplifting the known structure from Type IIB orientifolds.

Here we have been very explicit by working out the structure of two still very simple but non-toroidal Calabi-Yau orientifolds and their F-theory uplifts. Therefore, this work also puts the ideas and computational techniques about E3-brane instantons to a new level of sophistication in that it deals with genuine Calabi-Yau manifolds. From inspection of the results for the uncharged zero modes for these two examples, we were able to identify the line bundle cohomology of a vertical divisor on the Calabi-Yau fourfold with the equivariant line bundle cohomology of the corresponding divisor on the upstairs Calabi-Yau threefold.

We also analysed in detail the non-perturbative fate of perturbative $U(1)$ instantons. Only if the elliptic fiber does not receive the monodromy of an O7-plane, the $\ov\tau_{\dot\alpha}$ zero modes remain unlifted. This translates to the statement that $\ov\tau_{\dot\alpha}$ zero modes that survive non-perturbatively will must up as one-cycles of the M5-brane.

Uplifting the known results for charged matter instanton zero modes to F-theory, we delivered an inherent F-theory description for them. These matter zero modes arise on codimension one hypersurfaces in the base of the elliptically fibered vertical M5-brane divisor, over which the fiber degenerates according to the Kodaira classification. Yukawa type zero mode absorption interactions between two charged zero modes and a single matter field are supported on codimension two loci, i.e.~points, in the base of the M5-brane. The appearing F-theory zero modes and Yukawa type interactions generalize the known Type IIB results in that they also include possible enhancements to exceptional groups. A fair amount of analysis was done for the computation of matter zero modes arising from  the intersection of an instanton with the generic singular $I_1$ factor in the discriminant. Our result indicates that part of the Type IIB zero modes are non-perturbatively lifted when moving away from the Sen limit.

We have shown that the entire Hodge diamond of the M5-brane divisor in F-theory is determined by the geometry of the E3-brane in the upstairs Calabi-Yau threefold geometry and its intersection curves with the space-time filling D7-branes. The dictionary is summarized in Table \ref{tab_hodgediamond}. We gave a simple example supporting that this quite intriguing geometric relation continues to hold for more involved degenerations of the elliptic fibration.

Finally, we gave a Type IIB orientifold interpretation of how the three-cycles of the M5-brane contribute to the instanton one-loop determinant, and found that an E3-instanton has more potentially harmful moduli than meet the eye. From this we derived a sufficient criterion \eqref{thirdcriterion} to generate an uncharged, nowhere vanishing superpotential.

We focused on the fluxless case for simplicity, although it would be useful to generalize these results to the case with background four-form flux as considered in \cite{Tripathy:2005hv, Kallosh:2005gs,Bergshoeff:2005yp,Park:2005hj, Saulina:2005ve}, and further analyzed in \cite{Tsimpis:2007sx}.

Even though our methodology was  necessarily quite technical in nature, we would like to emphasize again that all  results obtained have direct consequences for the low energy effective action of F-theory compactifications on Calabi-Yau fourfolds. For an instanton to contribute to a certain coupling in the action, all its fermionic zero modes need to be absorbed. This implies in particular, that M5-brane instantons with charged matter zero modes can only contribute to those terms in the superpotential with the appropriate number of charged matter fields.

\subsection*{Acknowledgement}

We gratefully acknowledge discussions with Volker Braun, Thomas Grimm, Max Kreuzer, Timo Weigand, Martijn Wijnholt and would like to thank Thorsten Rahn and Helmut Roschy for proofreading the manuscript. The work of A. C. is supported by a EURYI award of the European Science Foundation, and in part by the Excellence Cluster Universe, Garching.

\appendix

\section{Equivariant index theorems}\label{app:index}

Index theorems are very useful tools that allow one to compute analytic indices of operators in terms of topological invariants of a space. A prominent example of a useful index theorem is formula for the Euler number of a complex n-manifold $M$ in terms of the top Chern class of its holomorphic tangent bundle $TM$:
\beq\label{noneqeuler}
  \chi(M) \equiv \sum_i^{2\,n} (-)^i\,b^i(M) = \int_M c_n(TM)\,,
\eeq
where the Betti numbers $b^i$ are the analytic indices of the de Rham exterior derivative.

Another important example is the Riemann-Roch theorem. Given a holomorphic vector bundle $V$ over $M$, one can define an analytic index for the holomorphic covariant derivative:
\beq
  \bar\partial_V = \bar\partial+A^{0,1}_V\, ,
\eeq
where $A^{0,1}_V$ is the $(0,1)$ part of the connection on $V$. Defining cohomology groups $H^i(M, V)$, which can be thought of as ``$V$-section valued $(0,i)$-forms'', the Riemann-Roch theorem gives us the relation:
\beq\label{noneqriemannroch}
  \chi(M, V) = \sum_i^{n} (-)^i\,{\rm dim} H^i(M, V) = \int_M {\rm ch}(V) \, {\rm Td}(TM)\,,
\eeq
where ${\rm ch}(V)$ is the Chern character of $V$, and the Todd class ${\rm Td}(TM)$ can be expanded as
\beq
  {\rm Td}(TM) = 1+\tfrac{1}{2}\,c_1+\tfrac{1}{12}\,(c_1^2+c_2)+\tfrac{1}{24}\,c_1\,c_3\,.
\eeq
The most useful application of this formula comes about, when one knows by other means that the groups vanish for $i>0$, in which case one is only left with a formula that computes the dimension of $H^0(M,V)$, i.e.~it counts the number of holomorphic sections of $V$.

There are, however, useful generalizations to index theorems that apply to situations where the manifold admits some sort of group action on it. More specifically, in the framework of orientifold compactifications, where we have a $\IZ_2$-involution $\sigma$ acting on our threefold $X$, one can define how the involution acts on the various cohomology groups of interest, and decompose the groups according to the $\IZ_2$-grading:
\beq
  H^i = H^i_+\oplus H^i_-\,.
\eeq
Similarly, one can generalize the concept of index, to a $\IZ_2$-\emph{equivariant}-index as follows:
\beq
  \sum_i^n (-)^i\,({\rm dim}H^i_+-{\rm dim}H^i_-)\,.
\eeq

The so-called equivariant index theorems allow us to compute such indices in terms of characteristic classes of the underlying space $M$, and the subspace $M^\sigma$ of fixed points under the involutions. We will now state two such useful theorems, which generalize the two theorems in \eqref{noneqeuler} and \eqref{noneqriemannroch}.

\subsection{Lefschetz fixed point theorem}

We begin by stating the \emph{Lefschetz fixed point theorem}, which is the equivariant generalization to \eqref{noneqeuler} applied to $\IZ_2$-involutions. See \cite{Distler:1987ee} and \cite{Brunner:2003zm} for more details. Let $M$ be a manifold of real dimension $m$, and $\sigma$ an involution with a fixed-point set $M^\sigma \subset M$. Defining an induced action on the De Rham cohomology of $M$, $H^*(M)$, such that $H^i = H^i_+\oplus H^i_-$, then we have
\beq
  L(\sigma, M) \equiv \sum_i^{\,m} (-)^i (b^i_+-b^i_-) = \chi(M^\sigma)\,,
\eeq
where the right-hand side is simply the Euler number of $M^\sigma$. This index is known as the \emph{Leftschetz number} of $\sigma$.

There is a very useful theorem that states the following equality:
\beq
  \bal
    \chi(M/\sigma) &{}= \tfrac{1}{2}\,\big(L(\sigma, M) + \chi(M)\big)\,,\\
    &{}= \sum_i^{\,m} (-)^i (b^i_+) 
  \eal
\eeq
In other words, Euler number of the $\IZ_2$-orbifolded space is given by the average of the Lefschetz number and the Euler number of $M$.

\subsection{Holomorphic Lefschetz theorem}

We will now state the equivariant generalization to the Riemann-Roch theorem \eqref{noneqriemannroch} for $\IZ_2$-involutions.

Let $M$ be a complex $n$-dimensional manifold, and $\sigma$ an involution acting on it with a fixed point set $M^\sigma$. Let $\pi: V \mapsto M$ be a holomorphic vector bundle on $M$, for which the action of $\sigma$ lifts to an action $\sigma^*$ on $V$, such that $\pi \circ \sigma^* = \sigma$. Then $\sigma^*$ acts on a section $s(x)$ of $V$ as follows:
\beq
  \sigma^*(s(x)) = \rho(\sigma)_V \times s(\sigma(x))\,
\eeq
where $\rho(\sigma)_V$ is some representation of $\IZ_2$ acting on $V$.

We may now decompose $H^*(M, V) = H^*_+(M, V) \oplus H^*_-(M, V)$, and define the holomorphic Lefschetz number as 
\beq
  \chi^{\sigma}(M, V) \equiv \sum_i^n (-)^i\,({\rm dim}H^i_+-{\rm dim}H^i_-)\,.
\eeq
We first state the holomorphic Lefschetz theorem, and then explain the notation:
\beq
  \chi^{\sigma}(M, V) = \int_{M^\sigma}\frac{\,{\rm ch_\sigma(V)}\, {\rm Td}(TM^\sigma)}{{\rm ch_\sigma}(\Lambda_{-1} \bar N_{M^\sigma})}\,.
\eeq
Here, $\Lambda_{-1} \bar N_{M^\sigma} = \sum_i^n \Lambda^i \bar N_{M^\sigma}$ is a formal alternating sum of exterior products of $\bar N_{M^\sigma}$, the complex conjugate of the normal bundle of $M^\sigma \subset M$.

The equivariant Chern character ${\rm ch_\sigma}(V)$ is defined by representing $V$ as a sum of line bundles, $L_i$, that are eigenbundles of $\sigma^*$, $V = L_1 \oplus \ldots \oplus L_k$. In other words, one decomposes $\rho_V(\sigma^*)$ into irreducible representations. Then, the definition of the character is the following:
\beq
  {\rm ch_\sigma}(V) = \sum_{j=1}^k \rho_j {\rm ch}(L_j)\, \quad {\rm where} \quad \rho_j = \pm 1\,.
\eeq

To demystify this relation a bit, let us restrict to the case where $M$ is a complex surface, and $M^\sigma$ is an isolated complex curve. Using the fact that, almost by definition, $\sigma^*(N_{M^\sigma}) = -N_{M^\sigma}$, we can write:
\beq
  \bal
    {\rm ch_\sigma}(\Lambda_{-1} \bar N_{M^\sigma}) &{}= {\rm ch_\sigma}(\cO -
    \bar N_{M^\sigma}) \\
    &{}= 1-{\rm ch_\sigma}(\bar N_{M^\sigma}) = 1+{\rm ch}(\bar N_{M^\sigma})\,.
  \eal
\eeq
With a bit of algebra, one arrives at the following formula:
\beq
  \chi^{\sigma}(M, V) =\int_{M^\sigma} \tfrac{1}{2}\,{\rm ch_\sigma}(V)\,\left(1+\tfrac{1}{2}\,c_1(M^\sigma)+\tfrac{1}{2}\,c_1(N_{M^\sigma})\right)\,.
\eeq
For trivial $V$, this gives us the following simple relation:
\beq
  \bal
    \chi^{\sigma}(M, \cO) &{}= \sum_i^2 (-)^i (h_+^{(0,i)}-h_-^{(0,i)})\\
                &{}= \tfrac{1}{4}\,\left(\chi(M^\sigma)+M^{\sigma} \cdot M^{\sigma}\right)\,,
  \eal
\eeq
where the last term is the self-intersection number of $M^\sigma$ in $M$.

For the case where $M$ is a divisor in a Calabi-Yau threefold $X$, the relation can be simplified further to the following:
\beq
  \chi^{\sigma}(M, \cO) = -\tfrac{1}{4}\,\int_{M^\sigma} [M]\,,
\eeq
where $[M]$ is meant as the element in $H^2(X)$, Poincar\'e dual to $M$. 

Note that we can extract the even and odd indices as follows:
\beq
  \bal
    \tfrac{1}{2}\,\left(\chi^{\sigma}(M, \cO)+\chi(M, \cO)\right) &{}= \sum_{i=0}\,(-)^i\,h^{0,i}_+\,,\\
    \tfrac{1}{2}\,\left(\chi^{\sigma}(M, \cO)-\chi(M, \cO)\right) &{}= \sum_{i=0}\,(-)^i\,h^{0,i}_- \,.
  \eal
\eeq


\clearpage
\nocite{*}
\bibliography{rev27}
\bibliographystyle{utphys}

\end{document}